\documentclass[%
 aip,
rsi,%
 amsmath,amssymb,
reprint,%
]{revtex4-1}

\usepackage{graphicx}
\usepackage{dcolumn}
\usepackage{bm}
\usepackage{color}
\usepackage{siunitx}

\begin{document}


\title[Dynamics and Rheology of Ring-Linear Blend Semidilute Solutions in Extensional Flow: Modeling and Molecular Simulations]
{Dynamics and Rheology of Ring-Linear Blend Semidilute Solutions in Extensional Flow: Modeling and Molecular Simulations}

\author{Charles D. Young}
\affiliation{Department of Chemical and Biomolecular Engineering, University of Illinois at Urbana-Champaign, Urbana, Illinois 61801, USA}
\affiliation{Beckman Institute for Advanced Science and Technology, University of Illinois at Urbana-Champaign, Urbana, IL, 61801}
\author{Yuecheng Zhou}
\altaffiliation{Current address: Department of Chemistry, Stanford University, Stanford, California 94305, USA}
\affiliation{Department of Materials Science and Engineering, University of Illinois at Urbana-Champaign, Urbana, Illinois 61801, USA}
\affiliation{Beckman Institute for Advanced Science and Technology, University of Illinois at Urbana-Champaign, Urbana, IL, 61801}
\author{Charles M. Schroeder}
\affiliation{Department of Materials Science and Engineering, University of Illinois at Urbana-Champaign, Urbana, Illinois 61801, USA}
\affiliation{Department of Chemical and Biomolecular Engineering, University of Illinois at Urbana-Champaign, Urbana, Illinois 61801, USA}
\affiliation{Beckman Institute for Advanced Science and Technology, University of Illinois at Urbana-Champaign, Urbana, IL, 61801}
\author{Charles E. Sing}
 \email{cesing@illinois.edu}
\affiliation{Department of Chemical and Biomolecular Engineering, University of Illinois at Urbana-Champaign, Urbana, Illinois 61801, USA}
\affiliation{Beckman Institute for Advanced Science and Technology, University of Illinois at Urbana-Champaign, Urbana, IL, 61801}

\date{\today}

\begin{abstract}

We use Brownian dynamics (BD) simulations and single molecule experiments to investigate the influence of topological constraints and hydrodynamic interactions on the dynamics and rheology of solutions of ring-linear polymer blends at the overlap concentration. We find agreement between simulation and experiment in that rings in solution blends exhibit large conformational fluctuations, including extension overshoots in the startup of flow and tumbling and tank-treading at steady state. Ring polymer fluctuations increase with blend fraction of linear polymers and are peaked at a ring Weissenberg number $\textrm{Wi}_R \approx 1.5$. On the contrary, linear and ring polymers in pure solutions show a peak in fluctuations at the critical coil-stretch Weissenberg number $\textrm{Wi} = 0.5$. BD simulations show that extension overshoots on startup of flow are due to flow-induced intermolecular ring-linear polymer hooks, whereas fluctuations at steady state are dominated by intermolecular hydrodynamic interactions (HI). This is supported by simulations of bidisperse linear polymer solution blends, which show similar trends in conformational dynamics between rings and linear polymers with a matched contour length. Compared to BD simulations, single molecule experiments show quantitatively larger fluctuations, which could arise because experiments are performed on higher molecular weight polymers with stronger topological constraints. To this end, we have advanced the understanding of the effects of topological interactions and intermolecular HI on the dynamics of semidilute ring-linear polymer blend solutions.
\end{abstract}

\maketitle

\section{\label{sec:Intro}Introduction}

The dynamics of ring polymers are of broad interest to fundamental polymer physics, \cite{rubinstein1986dynamics, mcleish2002polymers} technological applications, \cite{kaitz2013end, lloyd2018fully} and biological systems. \cite{taanman1999mitochondrial} In the context of polymer physics, ring polymers are a model for understanding the effect of chain architecture. The dynamics and rheology of linear polymer solutions at varying concentration have been widely studied. \cite{doi1988theory,rubinstein2003polymer} In the dilute limit, dynamics are governed by intramolecular excluded volume (EV) and hydrodynamic interactions (HI). Polymer solutions enter the semidilute regime above the overlap concentration $c^* \approx M/(N_A R_g^3)$, where $M$ is the polymer molecular weight, $N_A$ is Avogadro's number, and $R_g$ is the dilute polymer radius of gyration. In semidilute solutions, polymer dynamics are determined by both intra- and intermolecular EV and HI. Moreover, in melts EV and HI are fully screened, and polymer dynamics are well described by either the Rouse model for unentangled polymers or the tube-reptation model for entangled polymers. The introduction of an end-free constraint in ring polymers changes the conformational dynamics and rheology of polymers in all three concentration regimes qualitatively and quantitatively. 

In melts, polymer rheology can largely be understood on the basis of topological constraints. Linear polymer melts relax stress by reptation of free ends along the polymer backbone, leading to a rubbery plateau modulus. \cite{doi1988theory,rubinstein2003polymer} However, in ring polymer melts the end-free constraint significantly suppresses entanglements and drives rings toward globular conformations, resulting in a significantly lower zero-shear viscosity as compared to linear polymer melts and the absence of a rubbery plateau modulus. \cite{kapnistos2008unexpected, halverson2012rheology, ge2016self} These results are only valid for pure ring melts, which remain challenging to synthesize. \cite{doi2015melt,pasquino2013viscosity} The introduction of even trace linear contaminants results in significantly higher zero-shear viscosity and the return of a rubbery plateau modulus. \cite{kapnistos2008unexpected,halverson2012rheology} These rheological features are intrinsically connected to the polymer dynamics. Molecular dynamics (MD) simulations show long-lived relaxation modes in ring-linear polymer melt blends associated with the threading of linear polymers through rings. \cite{tsalikis2014threading} MD simulations also show a maximum in zero-shear viscosity as a function of blend ratio at $\phi_{linear} = 0.5$ (where $\phi_{linear}$ is the linear polymer weight fraction) due to ring-linear threading. \cite{halverson2012rheology}

Polymer processing generally involves strong flows, where the coupling of architecture and deformation complicate the idea of suppressed entanglements in pure ring melts at equilibrium. Filament stretching rheometry has shown that ring polymer melts in uniaxial extensional flow thicken at unexpectedly low strain rates, \cite{huang2019unexpected} a phenomenon which has been shown by MD simulations to be caused by topological linking. \cite{oconnor2020topological} Ring linking has also been observed in melts under shear,  although the effects on shear viscosity are less pronounced than in extension. \cite{halverson2012rheology, tsalikis2016analysis, tsamopoulos2019shear, jeong2020intrinsic} The effect of blend ratio is also relevant in flow. MD simulations show the shear viscosity is independent of blend ratio when melts are exposed to high shear rates, \cite{halverson2012rheology} whereas in extension threading-unthreading transitions lead to stress overshoots. \cite{borger2020threading}

In contrast to melts, topological constraints are weaker in solutions;  \cite{doi1988theory, rubinstein2003polymer} the coupling of HI and EV to polymer architecture becomes important, particularly in strong flows. \cite{mai2018stretching, schroeder2018single} In the dilute limit, equilibrium ring polymer dynamics and conformations have been widely studied by single molecule experiments, \cite{robertson2006diffusion} theory, \cite{jagodzinski1992universal} and simulation. \cite{hegde2011conformation, narros2013effects} In planar extensional flow, ring polymers exhibit a delayed coil-stretch transition due to intramolecular hydrodynamics which drive the ring to stretch in the flow-neutral direction into an open loop conformation. \cite{li2015ends,hsiao2016ring} In shear flow, rings undergo tumbling as observed in linear polymers, \cite{schroeder2005characteristic, tu2020direct} as well as tank-treading dynamics as seen in vesicles. \cite{chen2013tumbling,liebetreu2018trefoil} In the more general case of mixed flows, rings exhibit both tumbling and stretching behavior. \cite{young2019ring} Overall, rings in dilute solution demonstrate unique conformational dynamics and quantitatively different rheological responses as compared to linear polymers.

Semidilute polymer solutions incorporate the physics of both the melt and dilute regimes, namely entanglement or topological constraints and intra and inter-molecular interactions due to excluded volume and solvent-mediated hydrodynamics. \cite{doi1988theory,rubinstein2003polymer} At equilibrium, the influence of chain architecture in pure solution is primarily quantitative, as ring and linear polymers demonstrate similar scaling of dynamic and static properties with concentration and molecular weight. \cite{tsalikis2020conformation} However, blends of ring and linear polymers in semdilute solution can display qualitatively different dynamics, particularly above the entanglement concentration $c_e$ as traditionally defined for linear polymers. \cite{rubinstein2003polymer} Single molecule studies on the diffusion of trace rings in a background of semidilute linear polymers have shown a scaling with concentration lower than that of the pure linear solution prediction, suggesting ring-linear threading inhibits ring diffusion in a manner not described by reptation theories. \cite{robertson2007self} Further studies found that as the blend fraction of ring polymers decreased from a primarily ring to primarily linear polymer solutions, the diffusion of ring polymers decreased markedly. \cite{chapman2012complex} Monte Carlo simulations found good qualitative agreement with this trend, attributing the slower dynamics to ring-linear threading. \cite{chapman2012complex} This is supported by MD melt simulations, which show that ring-ring threadings are significantly less probable and have a weaker effect on ring dynamics as compared to ring-linear threads. \cite{tsalikis2016analysis} Notably, slow ring dynamics are most apparent above the entanglement concentration $c_e$, which for 48.5 kbp $\lambda$-DNA is related to the overlap concentration as $c_e \approx 3 c^*$. \cite{zhou2018entangled, pan2014scaling} At lower concentrations, the dependence on blend fraction is less pronounced. \cite{chapman2012complex}

The flow dynamics of semidilute solutions of pure rings or ring-linear polymer blends are still not well understood. In particular, it is not clear if the application of flow introduces topological constraints or strong solvent-mediated HI which are absent at equilibrium. Recently, Zhou et al. performed single molecule experiments on ring polymers trapped at the stagnation point of planar extensional flow in background solutions of linear polymers at $0.025 c^*$, $0.1 c^*$, and $1 c^*$. \cite{zhou2019effect} At $0.025 c^*$, ring polymer conformational fluctuations were peaked at $\textrm{Wi}_R \approx 0.5$, where $\textrm{Wi}_R$ is the dimensionless flow strength on trace ring polymers defined as $\textrm{Wi}_R = \dot{\epsilon} \tau_R$, $\dot{\epsilon}$ is the strain rate, and $\tau_R$ is the longest relaxation time of the trace ring polymer. This is in agreement with the expectation that conformational fluctuations are largest at the critical coil-stretch transition flow rate, as shown in previous experiments and simulations on semidulte linear polymer solutions, \cite{hsiao2017direct, sasmal2017parameter, stoltz2006concentration, young2019simulation} and dilute linear and ring polymer solutions. \cite{li2015ends, hsiao2016ring, young2019ring} At $0.1 c^*$, however, the maximum in conformational fluctuations was shifted to $\textrm{Wi}_R \approx 0.9$, with only a weak decrease thereafter. Approaching the semidilute regime at $1 c^*$, fluctuations increased up to $\textrm{Wi}_R \approx 1.5$ and then plateaued. These results raise questions regarding the nature of intermolecular interactions in non-dilute ring-linear polymer blends.

Single molecule experiments and Brownian dynamics (BD) simulations of linear semidilute polymer solutions have already suggested plausible explanations. At $1 c^*$, slow-stretching and fast-stretching end-coiled sub-populations emerge upon the startup of planar extensional flow, leading to broader conformational distributions as compared to dilute solutions. \cite{hsiao2017direct,sasmal2017parameter} The authors suggested this could be due to flow induced intermolecular hooks which form between folded and end-coiled conformations. Another BD simulation study at similar conditions found both broader conformational distributions at $1 c^*$ and transient intermolecular hooks, although hooks were rarely found at $1 c^*$. \cite{young2019simulation} Simulations also revealed a population of linear polymers which interconverted between coiled and stretched states at moderate flow rates $\textrm{Wi}_L \approx 1.5$, whereas linear polymers in dilute solution remained fully stretched. \cite{schroeder2004effect,young2019simulation} The slow stretching dynamics and retraction were assigned to intermolecular hydrodynamic interactions (HI), but a detailed mechanism was not proposed.

In this work, we utilize BD simulations to study the influence of planar extensional flow and blend composition in semidilute solutions of ring-linear polymer blends. We consider solutions at the overlap concentration $c^*$ while increasing flow rate through the coil-stretch transition, $\textrm{Wi}_R = 0.4 - 3.2$, and varying the blend composition from a pure ring polymer solution to a trace ring in a linear polymer background solution. We compare directly to single molecule experiments detailed in a companion article. \cite{zhou2021dynamics} The experiments reveal unexpected trends in ring dynamics with blend ratio and strain rate, which are reproduced in simulation and explained on the basis of intermolecular interactions.

 The article is organized as follows: In Sec. \ref{sec:Simulation Method} we describe the simulation method and experimental procedure. In Sec. \ref{sec:Results} we quantify the conformational dynamics and solution rheology. Ring polymers are found to exhibit large conformational fluctuations which are sensitive to the blend ratio. We further investigate the origin of ring conformational fluctuations in Sec \ref{sec:Discussion}. We find that intermolecular ring-linear polymer hooks lead to overshoots in ring extension on startup of flow, and intermolecular HI lead to large fluctuations in extension at steady state. Finally, we summarize our results and highlight topics for future study in Sec. \ref{sec:Conclusions}.

\section{\label{sec:Simulation Method}Methods}

\subsection{\label{sec:goveqn}Simulation Governing Equations}

We perform BD simulations of semidilute ring-linear polymer blend solutions in a planar extensional flow. The simulations consist of $n_{R}$ ring polymers and $n_{L}$ linear polymers each with the same number of coarse-grained beads per chain $N_R = N_L = 150$. The total number of beads is then $N = n_R N_R + n_L N_L$. The position $\bm{r}_{i}$ of a bead $i$ is updated according to the Langevin equation
\begin{equation}
    \frac{d\tilde{\bm{r}}_{i}}{d\tilde{t}} = \tilde{\bm{\kappa}} \cdot \tilde{\bm{r}}_{i} -\sum_{j} \tilde{\textbf{D}}_{ij} \nabla_{\tilde{\bm{r}}_{j}}(\tilde{U}) + \tilde{\bm{\xi}}_{i}
\end{equation}
Tildes denote dimensionless quantities. Positions are normalized by the bead radius ($\tilde{\bm{r}}=\bm{r}/a$), energies by the thermal energy $k_{B}T$ ($\tilde{U}=U/(k_{B}T)$), times by the single-bead diffusion time ($\tilde{t}=t/\tau_{0}$, where $\tau_{0}=6\pi \eta_s a^{3}/(k_{B}T)$ and $\eta_s$ is the solvent viscosity), and the diffusion tensor by the drag coefficient of the spherical polymer beads ($\tilde{\textbf{D}}_{ij}= \textbf{D}_{ij}(6\pi\eta_s a/(k_B T))$). Polymer beads experience flow via the block diagonal tensor $\tilde{\bm{\kappa}}$, which has $3 \times 3$ diagonal blocks given by the solvent velocity gradient tensor $(\nabla \tilde{\textbf{v}})^T$. For planar extensional flow,
\begin{equation}
    \nabla \tilde{\textbf{v}} = \begin{pmatrix} \tilde{\dot{\epsilon}} & 0 & 0 \\ 0 & -\tilde{\dot{\epsilon}} & 0 \\ 0 & 0 & 0 \end{pmatrix}
\end{equation}
where $\tilde{\dot{\epsilon}} = \dot{\epsilon} \tau_{0}$ is the dimensionless strain rate. Beads interact via a potential $\tilde{U} = \tilde{U}^{B} + \tilde{U}^{EV}$ consisting of bonded and excluded volume contributions. We use a finitely extensible non-linear elastic (FENE) spring force for connectivity
\begin{equation}
    \tilde{U}^{B} = -0.5 \tilde{k}_{s} \tilde{r}_{max}^{2} \textrm{ln} \left[1-\left( \frac{\tilde{r}_{ij}}{\tilde{r}_{max}} \right)^{2} \right]
\end{equation}
where $\tilde{k}_{s}=30 \tilde{u}/\tilde{\sigma}^{2}$ is the spring constant, $\tilde{u}=1.0$ gives the strength of EV interactions, and $\tilde{\sigma}=2$ is the diameter of a bead. The maximum extension of a spring is $\tilde{r}_{max}=1.5\tilde{\sigma}$, and $\tilde{r}_{ij}$ is the distance between two connected beads. Excluded volume interactions are modeled by a shifted, truncated, purely repulsive Lennard-Jones potential known as the Weeks-Chandler-Andersen (WCA) potential \cite{weeks1971role}
\begin{equation}
    \label{EV}
    \tilde{U}^{EV} = 4\tilde{u} \left[ \left( \frac{\tilde{\sigma}}{\tilde{r}_{ij}} \right)^{12} - \left( \frac{\tilde{\sigma}}{\tilde{r}_{ij}} \right)^{6} + \frac{1}{4} \right ] \Theta(2^{1/6} \tilde{\sigma} - \tilde{r})
\end{equation}
where $\Theta(x)$ is the Heaviside function. This EV potential yields chain statistics representative of a good solvent. We find $\nu \approx 0.59$ from the Zimm scaling relation $\tau_{Z}\sim N^{3\nu}$ and relaxation time data from equilibrium single chain simulations, in agreement with the result for a polymer in good solvent. \cite{doi1988theory,rubinstein2003polymer} This model has been widely utilized to study polymer dynamics in solution and melt and has been shown to prevent chain crossings in simulations of entangled melts in extensional flow. \cite{kremer1990dynamics} 

Solvent-mediated HI and Stokes drag are included via the diffusion tensor, given here by the Rotne-Prager-Yamakawa (RPY) tensor, \cite{rotne1969variational, yamakawa1970transport}
\begin{equation}
	\label{RPY}
    \tilde{\textbf{D}}_{ij} = \small \begin{cases}
    	\bm{I}, & i=j\\
        \frac{3}{4\tilde{r}_{ij}}\left[\left(1+\frac{2}{3{\tilde{r}_{ij}}^2}\right)\bm{I}+\left(1-\frac{2}{{\tilde{r}_{ij}}^2}\right)\bm{\hat{r}}_{ij}\bm{\hat{r}}_{ij}\right], & i\neq j, {\tilde{r}_{ij}}\geq 2\\
        \left(1-\frac{9{\tilde{r}_{ij}}}{32}\right)\bm{I}+\frac{3{\tilde{r}_{ij}}}{32}\hat{\bm{r}}_{ij}\hat{\bm{r}}_{ij}, & i\neq j, {\tilde{r}_{ij}}\leq 2\\
    \end{cases}
\end{equation}
$\hat{\bm{r}}_{ij} = \tilde{\bm{r}}_{ij}/\tilde{r}_{ij}$ is a unit vector in the direction of $\tilde{\bm{r}}_{ij}=\tilde{\bm{r}}_{j}-\tilde{\bm{r}}_{i}$ and $\bm{I}$ is the identity matrix. The average and first moment of the Brownian noise $\tilde{\bm{\xi}}_{i}$ are given by the fluctuation-dissipation theorem as $\langle \tilde{\bm{\xi}}_{i}(t) \rangle = 0$ and $\langle \tilde{\bm{\xi}}_{i}(t) \tilde{\bm{\xi}}_{j}(t') \rangle =2\tilde{\textbf{D}}_{ij}\delta(t-t')$ respectively. Simulation implementation requires the decomposition of the diffusion tensor as $\tilde{\textbf{D}} = \textbf{BB}^{T}$ so that the Brownian noise can be computed via $\tilde{\bm{\xi}}_{i}=\sqrt{2}\textbf{B}_{ij} \bm{f}_{j}$, where $\bm{f}_{j}$ is a Gaussian random variable with mean 0 and variance $dt$. In traditional BD simulations, evaluation of the Brownian noise is a computational bottleneck, with the cost scaling as $O(N^2) - O(N^3)$ depending on the algorithm. \cite{ermak1978brownian,fixman1986construction,geyer2009n,ando2012krylov} To bypass this expense, we use the iterative conformational averaging (CA) method. \cite{miao2017iterative, young2018conformationally, young2019simulation} A brief description of the method and extension to the case of ring-linear polymer blends is given in the Supplemental Information, and a more detailed derivation and verification can be found in the authors' previous work. \cite{miao2017iterative, young2018conformationally, young2019simulation}

Polymers are simulated in an initially rectangular simulation cell of volume $\tilde{V} = \tilde{l}_x \tilde{l}_y \tilde{l}_z$. The initial cell dimensions in the extension and compression directions are $\tilde{l}_x$ and $\tilde{l}_y$ respectively which must be equal due to the use of Kraynik-Reinelt boundary conditions for deformation of the cell with the flow. \cite{kraynik1992extensional, todd1998nonequilibrium} We specify the cell size in the neutral direction $\tilde{l}_z$ to be smaller than $\tilde{l}_x$ and $\tilde{l}_y$ so that the cell dimension in the extension direction is larger. This reduces finite size effects arising from polymers interacting with their own periodic images. A similar approach has been used in simulations of polymer melts in planar extensional flow, which found that results from the rectangular cell simulation were in quantitative agreement with results from a cubic box simulation. \cite{sefiddashti2018configurational} The cell volume is determined by $\tilde{V}=N/\tilde{c}$, where $\tilde{c}$ is the polymer concentration. We set the concentration via the normalized value $\tilde{c}/\tilde{c}^{*}_L$, where $\tilde{c}^{*}_L=N_{L}/(4/3 \pi \langle \tilde{R}_{g0,L} \rangle ^{3})$ is the overlap concentration. This defines the overlap with respect to the dilute linear polymer radius of gyration $\langle \tilde{R}_{g0,L} \rangle$. We have adopted this definition for consistency with the single molecule experiments. \cite{zhou2019effect} As a result, the effective normalized concentration decreases with increasing ring polymer blend fraction due to the smaller radius of gyration of the ring. The difference between the equilibrium sizes of ring and linear polymers is considerable ($\langle \tilde{R}_{g0,L} \rangle = 19.5$ vs $\langle \tilde{R}_{g0,R} \rangle = 14.5$), suggesting a change in the effective concentration may be important. However, we have performed simulations of pure ring polymer solutions at $\tilde{c}^*_{R}$ based on the ring polymer radius of gyration and found the results to be nearly quantitatively consistent with those presented here for $f_R=1$, which use $\tilde{c}/\tilde{c}^*_L = 1.0, \tilde{c}/\tilde{c}^*_R = 0.4$.  All following references to the overlap concentration $c^*$ indicate the value for the pure linear solution $\tilde{c}^*_L$, and tildes are dropped because only the normalized concentration is used.

We consider solution blends at the overlap concentration $c^*$ for a range of linear polymer fractions $f_R = 0.02-1$ and flow rates $\textrm{Wi}_R = 0.4-3.2$. The ring polymer fraction controls the blend ratio and is defined as $f_R = n_R N_R / (n_R N_R + n_L N_L)$. We define the ring polymer Weissenberg number $\textrm{Wi}_R = \dot{\epsilon} \tau_{R}$, where $\tau_R$ is the longest ring linear polymer relaxation time. We consider the single exponential ring relaxation time at the relevant blend fraction, which is determined as described in Section \ref{sec:relaxation}. The number of polymers in the simulation box and the resulting box dimensions are given in Table \ref{table:simulationparameters}. Note that simulations at $f_R = 0.02$ use a smaller box size for computational efficiency, as only one ring polymer trajectory is gathered per simulation run. We have also performed $n_{run} = 3$ simulation runs at $f_R =0.01$ using the same larger box dimensions as for $f_R = 0.17-0.83$. We find the linear polymer dynamics agree quantitatively between the larger and smaller boxes, so to study ring polymer dynamics we use the latter. A smaller simulation box size is also used for $f_R = 1$ because the contour length of ring polymers is half that of linear polymers.

\begin{table*}[t]
	\centering
    \begin{tabular*}{\textwidth}{c @{\extracolsep{\fill}} cccccc}
    \hline\hline
    	$f_R$ & $n_R$ & $n_L$ & $\tilde{l}_x, \tilde{l}_y$ & $\tilde{l}_z$ & $n_{run}$\\
        \hline
        1 & 34 $-$ 64 & 0 & 108.3 $-$ 148.5 & 90.0 & 3\\
        0.83 & 61 $-$ 106 & 13 $-$ 22 & 151.1 $-$ 199.3 & 100.0 & 3\\
        0.5 & 37 $-$ 64 & 37 $-$ 64 & 151.1 $-$ 199.3 & 100.0 & 3\\
        0.17 & 13 $-$ 22 & 61 $-$ 106 & 151.1 $-$ 199.3 & 100.0 & 3 $-$ 5\\
        0.02 & 1 & 39 $-$ 62 & 124.6 $-$ 156.3 & 80.0 & 40\\
    \hline\hline
    \end{tabular*}
    \caption{Simulation parameters for each blend ration $f_R$. $n_R$ is the number of rings polymers. $n_L$ is the number of linear polymers. $\tilde{l}_x, \tilde{l}_y,$ and $\tilde{l}_z$ are the initial simulation box dimensions in the extension, compression, and neutral directions respectively. $n_{run}$ is the number of simulation runs. The low end of the range corresponds to simulations at a flow rate $\textrm{Wi}_R = 0.4$, and the high end corresponds to simulations at $\textrm{Wi}_R = 0.8,1.6,3.2$.}
	\label{table:simulationparameters}
\end{table*}

Polymer conformations are initialized following a procedure inspired by simulation of ring-linear polymer blend melts. \cite{halverson2012rheology} Rings are introduced as randomly oriented circular ellipses on a  cubic  lattice with spacing greater than the diameter of the rings. This ensures that rings are initially non-concatenated. The number of beads is generally greater than intended because the box is cubic and the lattice is filled. Rings are randomly removed until the target number of beads $N$ is reached. Rings are then relaxed from their circular conformations in a freely draining (FD) simulation for a duration $\tau_R^{FD}$ corresponding to the FD ring relaxation time determined from dilute solution FD simulations. At this point, the concentration is lower than intended due to the large initial lattice spacing, so we decrease the box size to the target dimensions. The ring polymers are further relaxed for $10 \tau_R^{FD}$ before $n_L$ rings are removed and replaced with random-walk non-overlapping linear polymers to reach the target blend ratio. Finally, the system is allowed to relax for another $10 \tau_L^{FD}$ corresponding to the dilute solution FD linear polymer relaxation time. 

By this procedure, we aim to reach an accurate equilibrium conformation for a ring-linear polymer blend solution. As discussed in Section \ref{sec:hooking}, the initial probability of a linear polymer threading a ring is significant to the transient dynamics on startup of flow. Unfortunately, simulation studies on the equilibrium conformations, dynamics, and threading probability of ring-linear polymer blends in solution are limited, and existing results focus on concentrations $c > c_e$. The polymer density at $c^*$ is relatively low ($\rho = N/V = 0.04 / \sigma^3$), so we assume that our procedure provides accurate equilibrium conformations and continue to out-of-equilibrium dynamics and rheology, which are the focus of this work.

Initialization is followed by a production run including flow and HI. Kraynik-Reinelt boundary conditions (KRBCs) \cite{kraynik1992extensional} are implemented such that the simulation cell deforms consistently with the applied flow. We follow the algorithm of Todd and Daivis,\cite{todd1998nonequilibrium} which allows for unrestricted strain accumulation. Hydrodynamics are accounted for using an Ewald sum, \cite{beenakker1986ewald, jain2012optimization} which overcomes the slow convergence of the RPY tensor by splitting the sum into exponentially decaying real space and reciprocal space parts. Excluded volume interactions are accelerated using a cell list generalized for homogeneous linear 3D flow. \cite{dobson2016cell} As the simulation progresses, the accumulated Hencky strain is given by the applied flow rate $\epsilon_H = \dot{\epsilon} t$. We simulate until a total strain $\epsilon_{tot} = 15-20$, after which flow is halted and the box remains in the conformation at the cessation flow time $t_{cess}=\epsilon_{tot}/\dot{\epsilon}$. We then simulate relaxation dynamics for $~10 \tau_{L}^{FD}$. The simulation is advanced by explicit Euler integration of the Langevin equation using a time step of $dt=5 \times 10^{-4} \tau_{0}$.

\subsection{Experimental Methodology}
To prepare ring-linear polymer blend solutions, small amounts of 45 kbp circular DNA molecules are first fluorescently labeled with an intercalating dye (YOYO-1, Molecular Probes, Thermo Fisher) with a dye-to-base pair ratio of 1:4 for $>$1 h in dark at room temperature. Trace amounts of fluorescently labeled 45 kbp DNA are then added to background solutions of unlabeled 45 kbp semidilute ring-linear DNA blends. Details regarding the preparation of 45 kbp circular DNA and semidilute ring-linear DNA blend solutions are described elsewhere. \cite{robertson2006diffusion, zhou2019effect, zhou2021dynamics}

Single molecule fluorescence microscopy and imaging is performed using an inverted epifluorescence microscope (IX71, Olympus) coupled to an electron-multiplying charge coupled device (EMCCD) camera (iXon, Andor Technology) as described in detail before. \cite{zhou2016expt, zhou2018entangled}. In brief, labeled DNA blend solutions are introduced into a PDMS-based microfluidic cross-slot with 300 $\mu$m in width and 100 $\mu$m in height. A 50 mW 488 nm laser directed through a 2.2 absorbance neutral density (N.D.) filter (Thorlabs, NJ, USA) is reflected by a 488 nm single edge dichroic mirror (ZT488rdc, Chroma) and used to illuminate the labeled DNA molecules. Fluorescence emission is collected by a 1.45 NA, 100$\times$ oil immersion objective lens (UPlanSApo, Olympus) followed by a 1.6$\times$ tube lens and a 525 nm single-band bandpass filter (FF03-525/50-25, Semrock) in the detection path. Fluorescence images are acquired by an Andor iXon EMCCD camera (512$\times$512 pixels, 16 $\mu$m pixel size) under frame transfer mode at a frame rate of 33 Hz (0.030 s$^{-1}$). Additional details regarding the experimental methods can be found in the companion article. \cite{zhou2021dynamics}

\subsection{Comparison between simulation and experiment\label{sec:method_diff}}

Simulation variables are matched to experimental conditions as possible, and generally we find good qualitative agreement. However, quantitative comparison remains challenging due to computational expense and availability of appropriate models. First, we use a flexible chain with a FENE-WCA force law, as compared to the wormlike force-extension behavior of DNA. Polymer stiffness influences the extensional rheology of polymer solutions. \cite{dinic2020flexibility} The FENE-WCA force law was used for convenient implementation of topological constraints on the length scale of an individual Kuhn segment. Generally, WLC models used in simulation are coarse-grained, \cite{marko1995stretching} which is not sufficient for capturing hooking behavior. WLC models on the level of an individual Kuhn segment have been developed, \cite{underhill2006alternative, saadat2016new} but they are challenging to implement with steep LJ excluded volume potentials to prevent spring crossings. Alternative algorithms for enforcing topological constraints such as spring-spring repulsions \cite{kumar2001brownian} or slip-links \cite{likhtman2005single, uneyama2012multi, ramirez2015multichain} may be useful, but we have not explored them here.  Ultimately, the goal of the simulations is to determine the effect of topological constraints and intermolecular HI on ring-linear blend dynamics, both of which are achieved by our model. We show that these are the origin of ring extension fluctuations observed in experiment, although the simulations cannot quantitatively reproduce experimental measurements. 

Second, simulations consider relatively short chains with $n_K \approx 83$ Kuhn segments per chain as compared to $n_K \approx 200$ for 45 kbp ring DNA and $\lambda$-DNA (48.5 kbp). Simulations are limited to short chains because of the size of the box, with the current $N_R = 150$ systems using $N = 19,200$. Higher molecular weight chains would require larger systems, which are intractable due to the $O(N^2)$ computational scaling of the CA method. BD algorithms with improved computational scalings of $O(N)$ \cite{fiore2017rapid} or $O(N \textrm{log} N)$ \cite{liu2014large,saadat2015matrix} may help in overcoming this limitation, but they have not been implemented in the CA method. Prior work has shown that even with the lower molecular weight, simulations still capture the same qualitative trends as experiments. \cite{hsiao2016ring, young2019simulation, patel2020nonmonotonic}

 Finally, simulations use periodic boundary conditions in a homogeneous unbounded flow. The flow kinematics in a cross-slot clearly differ, as previous work has shown asymmetric unsteady flow of synthetic flexible polymers which is sensitive to channel shape, \cite{arratia2006elastic, haward2016elastic, cruz2018characterization} channel aspect ratio, \cite{sousa2015purely, cruz2016influence} polymer concentration, and molecular weight. \cite{haward2016elastic, sousa2015purely} However, flow instabilities have not been reported for DNA solutions in cross-slot devices, \cite{hsiao2017direct, zhou2019effect} which is confirmed in the companion article to the current study. \cite{zhou2021dynamics} The higher persistence length of DNA may lead to lower solution elasticity, suppressing the instability entirely or delaying it to high Wi. Molecular simulations of polymers in a cross-slot microfluidic device may be able to resolve this issue, \cite{delong2014brownian} but available algorithms are generally limited to dilute solutions due to computational expense. \cite{zhao2017parallel} 

\section{\label{sec:Results}Results}

To investigate polymer dynamics, we primarily consider the fractional extension in the flow direction, $\Delta x/L$. In simulation, the extension is $\Delta x = \textrm{max}(x_i) - \textrm{min}(x_i)$. For linear polymers, the contour length is $L_L = (N_L - 1) r_{max}$, whereas for ring polymers $L_R = (N_R - 1) r_{max} / 2$ due to the closed loop constraint. In experiment, the extension can be directly visualized and normalized by the ring contour length $L_R = 10 \mu$m.

We also probe the solution rheology via the reduced extensional viscosity, in which we have normalized by the monomer concentration to account for the linear concentration dependence. For dimensional consistency, we introduce a reference concentration $c^*$
\begin{equation}
    \eta_r = \frac{\eta_p c^{*}}{\eta_s c}
\end{equation}
where $\eta_p$ is the polymer contribution to the extensional viscosity
\begin{equation}
    \eta_{p}=-\frac{\tau_{p,xx}-\tau_{p,yy}}{\dot{\epsilon}}
\end{equation}
and $\tau_{p,\alpha \beta}$ is the polymer contribution to the stress tensor determined by the Kirkwood formula\cite{doi1988theory}
\begin{equation}
    \tau_{p,\alpha \beta} = \frac{1}{V} \sum_{i}^{N} \sum_{j>i}^{N} r_{ij,\alpha} F_{ij,\beta}
\end{equation}
$F_{ij,\beta}$ is the conservative force between particles $i$ and $j$ in the $\beta$ direction.

\subsection{\label{sec:relaxation}Relaxation after flow cessation}

\begin{figure}[htb]
	\includegraphics[width=0.5\textwidth]{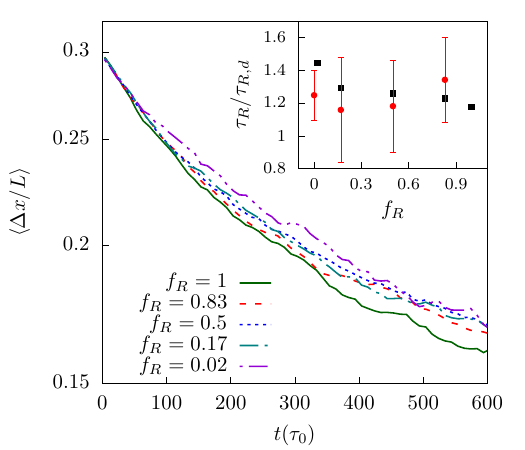}
    \caption{ Simulation  ensemble average fractional extension $\Delta \langle x / L \rangle$ relaxation from stretched steady state conformations at $\textrm{Wi}_R \approx 1.5$ after flow cessation for decreasing blend fraction of rings $f_R = 1$ (green solid line), $f_R = 0.83$ (dashed red line), $f_R = 0.5$ (dotted blue line), $f_R = 0.17$ (dash dot teal line), and $f_R = 0.02$ (dot dash violet line). Inset: Ring polymer relaxation time at $c^*$ normalized by the dilute limit value as a function blend ratio from simulation (black squares) and experiment (red circles).
    }
    \label{fig:rlx}
\end{figure}

The longest polymer relaxation time is determined by fitting a single exponential to the linear entropic regime \cite{schroeder2018single}  $\Delta x / L < 0.3$ after cessation of constant strain rate flow following an applied strain $\epsilon = 20$.  Fig. \ref{fig:rlx} shows the ensemble average relaxation for varying blend ratio determined from simulations.  In all cases we find a good fit to $\Delta x^2 = (\Delta x_0^2 - \Delta x_{\infty}^2)\textrm{exp}(-t/\tau_R) + \Delta x_{\infty}^2$, where $x_0/L = 0.3$ and $x_\infty/L \approx 0.13$ is the average extension at times $t > 5 \tau_R$ after flow cessation when the ring has fully relaxed. The resulting ring polymer relaxation time $\tau_R$ normalized by the dilute limit value $\tau_{R,d}$ is plotted versus blend ratio in the inset and compared to results from single molecule experiments. We find the relaxation time decreases weakly with blend fraction of rings $f_R$. Simulations agree with experiments to within the stochastic distribution of single molecule relaxation times.

\begin{figure*}[htb]
	\includegraphics[width=\textwidth]{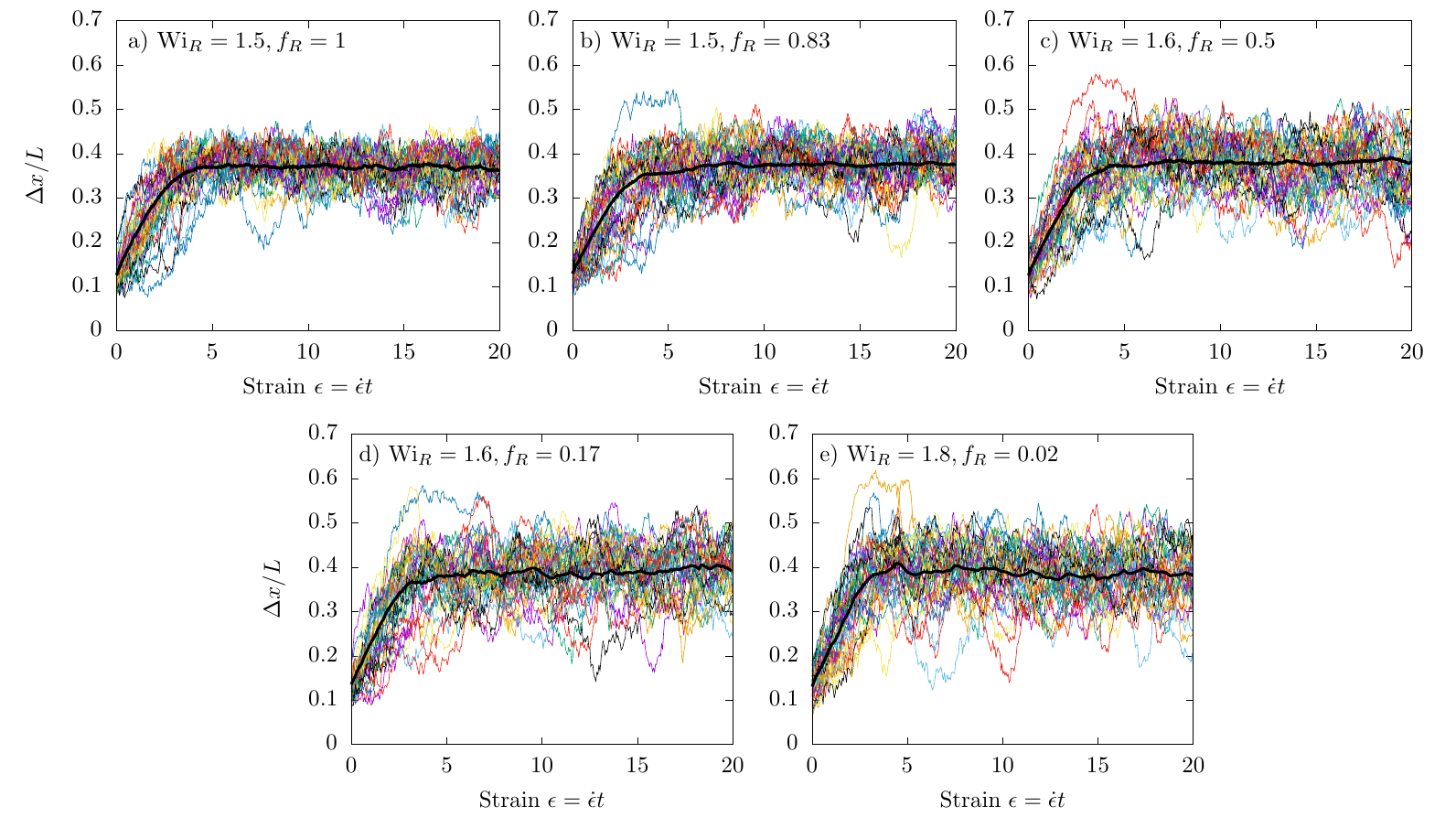}
    \caption{BD simulation results for individual (colored) and ensemble averaged (black) molecular trajectories of ring polymer fractional extension $\Delta x/L$ vs accumulated Hencky strain $\epsilon$ with decreasing ring polymer fraction of a) $f_R = 1$, b) $f_R = 0.83$, c) $f_R = 0.5$, d) $f_R = 0.17$ and e) $f_R = 0.02$ at a fixed flow strength of $\textrm{Wi}_R \approx 1.5$. Only 40 trajectories of the 100-200 molecule ensembles are shown for visual comparison with experiments. Rings in blend with linear polymers ($f_R < 1$) exhibit overshoots in extension on startup of flow and large fluctuations at steady state.
    }
    \label{fig:traj_vs_fL_sim}
\end{figure*}

We fit the linear polymer relaxation trajectories in the same way and find a similar dependence on blend fraction. Generally, for the same molecular weight, the linear polymer relaxation time $\tau_L$ is larger than $\tau_R$ ($\tau_L \approx 3.5 \tau_R$) because rings are topologically constrained such that they cannot satisfy the lowest order Rouse mode boundary condition. \cite{li2015ends, hsiao2016ring} We primarily refer to the Weissenberg number as defined using the ring polymer relaxation time $\textrm{Wi}_R$ to describe ring dynamics. However, when considering the solution average flow properties of blends, there are a spectrum of relaxation modes associated with the two components. Thus, we also report the linear polymer relaxation time when appropriate, such as in the solution flow modification for a trace ring in a semidilute linear background (Sec. \ref{sec:bulkflow}). It may be useful to determine a nominal relaxation time from the decay of solution average properties including the extensional viscosity and the birefringence, but in this study we consider only the polymer conformational relaxation time.

\subsection{Transient molecular conformations}

Next we investigate the transient conformations of ring polymers in startup and steady state planar extensional flow. In Fig. \ref{fig:traj_vs_fL_sim} we present BD simulation results at a fixed strain rate and decreasing blend fraction of rings $f_R = 0.02 - 1.00$. The ring polymer Weissenberg number $\textrm{Wi}_R \approx 1.5$ is approximately constant, with slight variations because the ring polymer relaxation time decreases with ring blend fraction.

For a pure ring solution, polymers stretch in the flow direction and then exhibit small fluctuations around the steady state average. The ensemble average extension reaches a constant at $\epsilon \approx 4-5$. This steady state is achieved faster than in the case of dilute or semidilute linear polymer solutions, consistent with previous experiments and simulations of dilute ring polymer solutions. \cite{li2015ends, hsiao2016ring}  The faster rate of stretching of ring polymers is ascribed to the lack of free ends, making the slow-stretching folded pathways observed for linear polymers less accessible to rings. Instead, rings stretch primarily via the faster dumbbell and half-dumbbell conformational pathways.  Once steady state is reached, the dynamics are consistent with previous semidilute linear polymer solution simulations. \cite{young2019simulation}

For blends of ring and linear polymers, we find markedly different behavior. While the ensemble average fractional still plateaus at $\epsilon \approx 4-5$, a sub-population of rings stretches significantly beyond the average to $\Delta x / L \approx 0.5-0.6$.   This behavior is most noticeable upon the startup of flow, although for majority linear polymer blends, rings can also reach highly stretched conformations $\Delta x / L > 0.5$ at steady state. Additionally, after the steady state average extension is reached, rings can retract back to equilibrium levels of extension $\Delta x /L \approx 0.1-0.2$. The extension of individual rings fluctuates significantly in time, consistent with the authors' previous \cite{zhou2019effect} and current experiments. \cite{zhou2021dynamics} These fluctuations grow in magnitude as the blend fraction of rings decreases, which we quantify in the following section. 

\begin{figure*}[htb]
	\includegraphics[width=\textwidth]{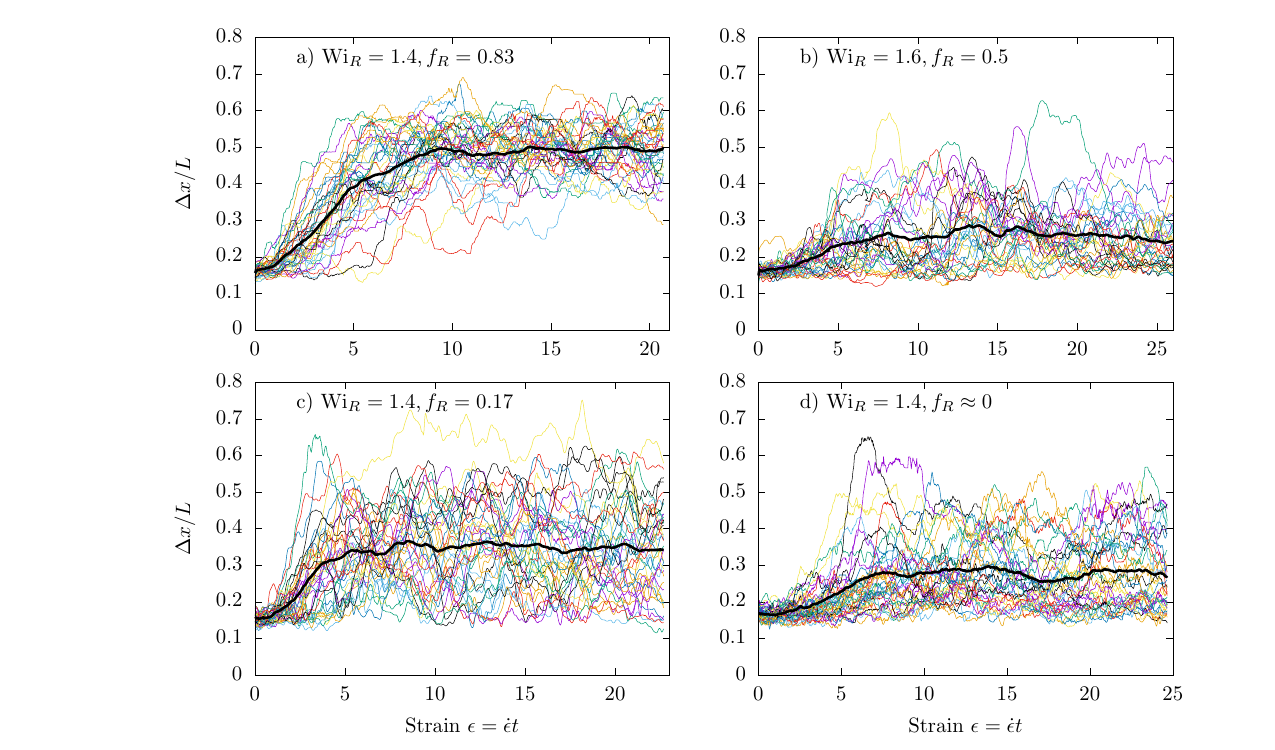}
    \caption{Single molecule experiment results for individual (colored) and ensemble averaged (black) molecular trajectories of ring polymer fractional extension $\Delta x/L$ vs accumulated Hencky strain $\epsilon$ with decreasing ring polymer fraction of a) $f_R = 0.83$, b) $f_R = 0.5$, c) $f_R = 0.17$ and d) $f_R \approx 0$ at a fixed flow strength $\textrm{Wi}_R \approx 1.5$. Experiments exhibit qualitative agreement with simulations in extension overshoots on startup of flow and fluctuations at steady state.}
    \label{fig:traj_vs_fL_expt}
\end{figure*}

We also present results from experiments at similar conditions in Fig. \ref{fig:traj_vs_fL_expt}, where $\textrm{Wi}_R \approx 1.5$ and $f_R \approx 0 - 0.83$. The majority ring polymer solution $f_R = 0.83$ is consistent with the $f_R = 1$ simulation results. Rings stretch to the steady state average and exhibit small fluctuations. As the blend fraction of rings decreases, large fluctuations emerge as seen in simulation. There are quantitative differences in the magnitude of fluctuations, which we ascribe to the differences between simulation and experiment discussed in Section \ref{sec:method_diff}. The experimental results are analyzed and discussed further in the companion paper. \cite{zhou2021dynamics} In particular, temporal correlations in extension and changes to the coil-stretch transition are investigated.

\subsection{\label{sec:avg_fluc}Average conformational fluctuations}

\begin{figure*}[htb]
	\includegraphics[width=\textwidth]{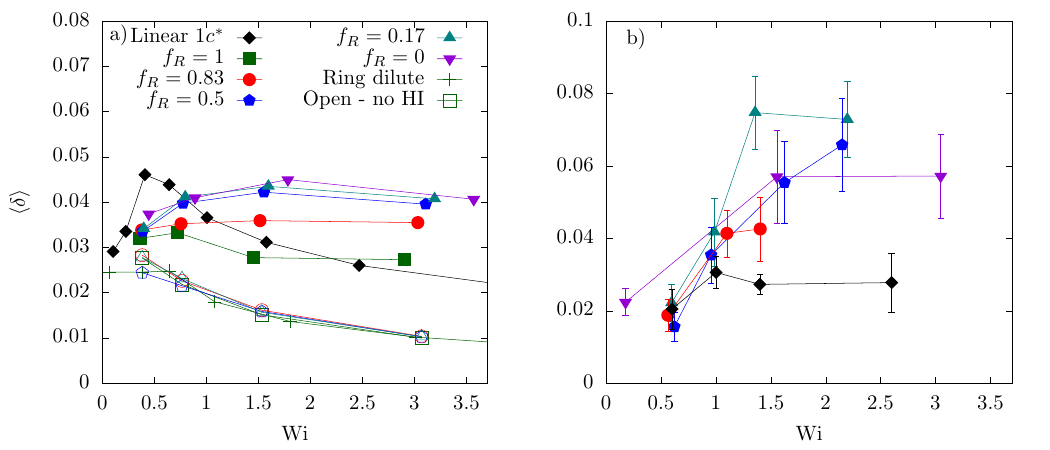}
    \caption{Ensemble average steady state fluctuation quantity $\langle \delta \rangle$ as a function of Wi for varying blend ratio as determined from a) simulation b) experiment. The Weissenberg number Wi is defined with the appropriate relaxation time for either a ring or linear polymer at the relevant concentration and blend ratio. Closed and open symbols in a) correspond to simulations with and without HI respectively.  Green crosses correspond to dilute ring polymer solutions. Open symbols are overlapping with green crosses, indicating intermolecular HI causes fluctuations in semidilute blends.  Conformational fluctuations of ring polymers in blend ($f_R < 1$) increase up to $\textrm{Wi}_R \approx 1.5$, in contrast to pure ring and pure linear solutions, where $\langle \delta \rangle$ is peaked at the coil-stretch transition flow rate $\textrm{Wi} = 0.5$.
    }
    \label{fig:fluc_avg}
\end{figure*}

We quantify the conformational fluctuations described above via the steady state ensemble average fluctuation quantity
\begin{equation}
    \langle \delta \rangle = \frac{\sum_{i=1}^{n} \sum_{\epsilon_{ss}}^{\epsilon_{cess}} \sqrt{(x_i(t)/L - \langle x_i/L \rangle)^2}}{n(\epsilon_{cess} - \epsilon_{ss})}
\end{equation}
where $n$ is the ensemble size, $\epsilon_{ss}$ is the accumulated strain at which the ensemble average fractional extension plateaus, and $\epsilon_{cess}$ corresponds to flow cessation. We thereby remove effects of initial transient stretching. This definition is consistent for both ring and linear polymers, with the linear steady state strain $\epsilon_{ss} \approx 8-9$, as compared to $\epsilon_{ss} \approx 4-5$ for the rings.

In Fig. \ref{fig:fluc_avg}, we report the fluctuation quantity as a function of Wi for a variety of architectures and blend ratios as determined from simulation and experiment. In all cases, the dimensionless flow rate Wi is determined using the relaxation time upon cessation of planar extensional flow at the relevant concentration and blend ratio. In addition to the simulations and experiments presented in this work, we have included results for the case of a pure linear solution at $c^*$ and a dilute ring polymer solution from previous work. \cite{young2019simulation, hsiao2017direct}

In simulation, the pure linear solution undergoes a maximum in the fluctuation quantity at $\textrm{Wi} \approx 0.5$, after which $\langle \delta \rangle $ decreases. This is expected because conformational fluctuations are largest at the coil-stretch transition. \cite{de1974coil} For dilute polymer solutions and semidilute solutions performed in the previous work, \cite{young2019simulation} we find consistent behavior with slight quantitative shifts (Supplementary information Fig. S1). The fluctuation quantity for pure linear solutions determined from experiment is consistently peaked at low Wi, although quantitatively smaller. 

Dilute ring polymer solutions exhibit suppressed conformational fluctuations as compared to pure linear polymer solutions. This is also expected due to the constrained conformations of the ring. In dilute solution, the majority of linear polymer fluctuations arise from end retraction, which are absent in the ring case. This trend appears to be consistent for semidilute pure ring polymer solutions. Fluctuations are larger than the dilute case due to intermolecular HI but are again peaked at low Wi, and both are lower than the linear polymer case. The semidilute ring polymer solution plateaus at high Wi and meets the pure linear solution result.

A distinct departure from the behavior of either pure linear or ring polymer solutions is observed in the case of ring polymers in ring-linear semidilute blends. For all blend ratios presented here, fluctuations are not peaked at $\textrm{Wi} \approx 0.5$. Instead, they increase up to $\textrm{Wi} \approx 1.5$, with a weak decrease thereafter. Furthermore, fluctuations increase as the blend ratio shifts towards linear chains from $f_R = 0.83$ to $f_R = 0.02$, confirming the visual observations of molecular trajectories in Figs. \ref{fig:traj_vs_fL_sim} and \ref{fig:traj_vs_fL_expt}. Simulation results show that most of this increase occurs from $f_R = 1$ to $f_R = 0.5$. As more linear chains are added, further increases in fluctuations are small. This is largely consistent with experiments, which also show increasing $\langle \delta \rangle$ with blend fraction of linear polymers.

To study the influence of HI, we perform `freely-draining' (FD) simulations which neglect HI for semidilute solutions at blend ratios $f_R = 1 - 0.17$. The relaxation time used to define $\textrm{Wi}_R$ is the Rouse relaxation time. Because the Rouse time is greater than the Zimm relaxation time obtained with HI, \cite{doi1988theory,rubinstein2003polymer} the strain rate is reduced to obtain the same values of $\textrm{Wi}_R$.  In the absence of HI, ring fluctuations are significantly suppressed, as shown by the open points in Fig. \ref{fig:fluc_avg}a, which are overlapping with the dilute ring fluctuations (green crosses). Additionally, there is no blend ratio dependence in semidilute FD simulations. The quantitative agreement between semidilute blends without HI and dilute solutions with HI suggests that there are minimal intermolcular interactions in FD simulations. This is consistent with previous BD simulations, which have shown FD simulations exhibit weak concentration dependence in planar extensional flow. \cite{stoltz2006concentration,young2019simulation} We provide a physical mechanism for enhanced ring conformational fluctuations on the basis of intermolecular HI in Sec. \ref{sec:himod}. 

As compared to the ring polymer component of the blends, the linear polymer component exhibits conformational fluctuations nearly quantitatively consistent with the pure linear semidilute solution case ( SI Fig. S1). There is a slight quantitative increase in fluctuations with linear polymer fraction, but the trend is weak compared to the ring polymer case. Thus to the knowledge of the authors, the overshoots in ring extension on startup of flow and the fluctuations at steady state appear unique to rings in semidilute blend with linear polymers. 

\subsection{Steady state conformations and rheology}

\begin{figure*}[htb]
	\includegraphics[width=\textwidth]{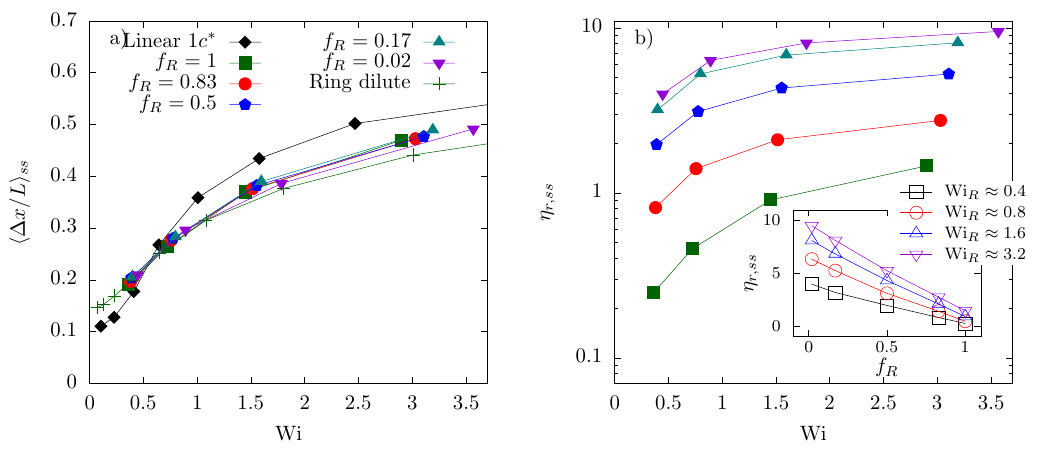}
    \caption{Ensemble average steady state a) fractional extension $\langle \Delta x / L \rangle_{ss}$ b) reduced extensional viscosity $\eta_{r,ss}$ as a function of $\textrm{Wi}_R$ for varying blend ratio as determined from simulation. Inset: Same data as b) but plotted versus blend ratio $f_R$ at approximately constant $\textrm{Wi}_R$. Ring polymer extension is nearly independent of blend ratio, but the extensional viscosity decreases with $f_R$ because linear polymers are more stretched than rings. Extensional viscosity varies linearly with blend ratio because the stress is dominated by the linear polymers.
    }
    \label{fig:ext_eta_avg}
\end{figure*}

The steady state conformations and extensional viscosity of dilute polymer solutions are well understood for linear and ring architectures. \cite{schroeder2004effect, li2015ends} Both undergo a transition from an equilibrium coil to a stretched conformation at $\textrm{Wi} \approx 0.5$, although the ring transition is more gradual due to intramolecular hydrodynamics. \cite{hsiao2016ring} In unentangled solutions, the extensional viscosity is dominated by polymer stretching. The large fluctuations in extension observed for ring polymers in semidilute blends motivate study of the effect of blend ratio on the ensemble average stretch and bulk viscosity. 

In Fig. \ref{fig:ext_eta_avg}a we plot the ensemble average steady state ring fractional extension after $\epsilon \approx 4-5$ versus the blend ratio $f_R$ and $\textrm{Wi}_R$. We also include results from pure linear solutions at $c^*$ for comparison. The data for the linear component of the blend is not shown because the linear polymer relaxation time is larger than the ring relaxation time, $\tau_L \approx 3.5\tau_R$. Thus the effective linear polymer Weissenberg number $\textrm{Wi}_L = \dot{\epsilon} \tau_{L}$ at the same strain rate is higher, and the linear chains are stretched for all strain rates presented here. We first observe that the more gradual coil-stretch transition found in dilute solution\cite{hsiao2017direct} persists in semidilute solution. This is expected because HI is nominally unscreened at the overlap concentration, and intramolecular HI drives ring extension in the neutral $z$ direction. \cite{hsiao2016ring} The open ring conformation is significant to both topological interactions and intermolecular HI, as shown in Section \ref{sec:Discussion}.

Considering the effect of blend ratio, the average extension curves collapse nearly quantitatively. This suggests that the dominant contribution to the average stretch is only the dimensionless flow strength $\textrm{Wi}_R$. In the following section we argue this is consistent with the large transient fluctuations. In particular, we show how intermolecular HI and topological interactions can drive instantaneous retraction and extension of the ring polymers. 

We consider the influence of blend ratio on steady reduced extensional viscosity $\eta_{r,ss}$ in Fig. \ref{fig:ext_eta_avg}b. As expected, viscosity increases with deceasing fraction of rings because the stress in unentangled polymer solutions is dominated by stretching. Linear polymers are more stretched than rings at the same strain rate, so as $f_R$ decreases, the linear chain contribution to the polymer stress dominates. When we plot viscosity against $f_R$ at constant strain rate (inset), we find a nearly linear relationship, suggesting the bulk viscosity is determined by simply mixing the linear and ring polymer contributions.  This is supported by plotting the ring and linear contributions to the extensional viscosity separately and normalizing by the number of molecules of that identity in solution (SI Fig. S2). 

\subsection{Conformational distributions}

\begin{figure*}[htb]
	\includegraphics[width=\textwidth]{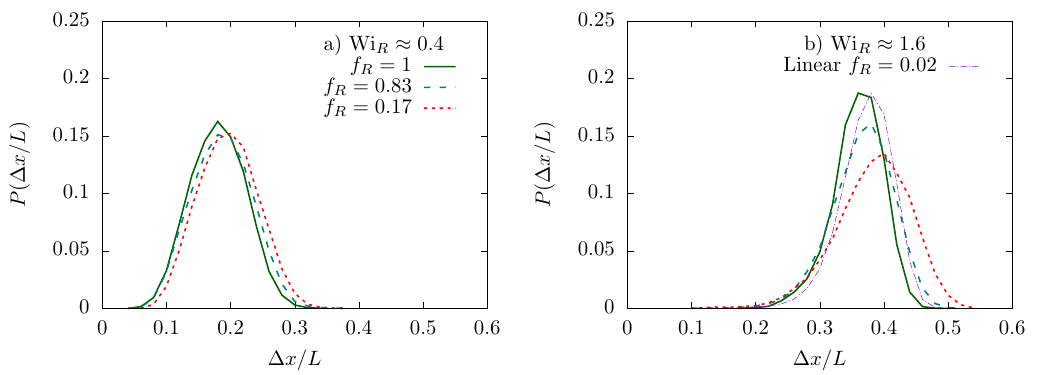}
    \caption{Steady state distributions of ring polymer fractional extension at blend fractions $f_R = 1$ (solid lines), $f_R = 0.83$ (dashed lines), and $f_R = 0.17$ (dotted lines) for increasing flow rate at a) $\textrm{Wi}_R \approx 0.4$, b) $\textrm{Wi}_R \approx 1.6$. For comparison, we include linear polymer distributions (dash-dot lines) in a nearly pure linear polymer solution ($f_R = 0.02$) at similar flow rates with respect to the linear polymer relaxation time at b) $\textrm{Wi}_L \approx 1.4$. Distributions for rings in blend with linear polymers ($f_R < 1$) are broader than for rings in pure solution and linear polymers at all blend ratios.
    }
    \label{fig:dss}
\end{figure*}

We conclude our characterization of ring polymer conformations with probably distributions of fractional extension, $P(\Delta x/L)$, in steady and startup flow. Conformational distributions have been widely used to quantify molecular individualism in dilute solution. Currently, we investigate the effect of intermolecular interactions on molecular individualism.

In Fig. \ref{fig:dss}, we plot the steady state distributions of ring polymer extension for approximately fixed $\textrm{Wi}_R$ and varying blend ratio $f_R = 0.17, 0.83, 1.00$. Results for blend fractions $f_R = 0.02, 0.50$ exhibit near quantitative matching with $f_R = 0.17$ and have been omitted for visual clarity. At low flow rates $\textrm{Wi}_R \approx 0.4$, rings are relatively unperturbed from their equilibrium conformations, and the distribution is relatively insensitive to the blend ratio. The linear chains are moderately stretched to $\Delta x / L \approx 0.4$ because the effective linear flow strength is $\textrm{Wi}_L \approx 1.4$. However, given the small change in ring conformations, it appears that interactions with linear polymers do not affect ring dynamics in weak flow.

When exposed to stronger flows, $\textrm{Wi}_R = 0.8-3.2$, rings are stretched from their equilibrium conformations. Notably, distributions broaden with decreasing fraction of ring polymers $f_R$ for all flow rates $\textrm{Wi}_R > 0.5$. For comparison, we include distributions for linear polymers in a nearly pure linear blend ($f_R = 0.02$) at matched effective flow strength $\textrm{Wi}_R \approx \textrm{Wi}_L$. The comparison clearly shows that the conformational distributions of ring polymers in pure ring solutions agree with those of linear polymers. Specifically, distributions become narrower with increasing flow strength. This is not the case for rings in blends with linear polymers, where distributions are broadest at $\textrm{Wi}_R \approx 1.6$, and broader than the pure polymer solution cases at $\textrm{Wi}_R \approx 0.8$ and $\textrm{Wi}_R \approx 3.2$. While the low extension tail is similar for all architectures and blend ratios, ring polymers in blends exhibit a high extension tail that becomes more pronounced with decreasing $f_R$.

Transient conformational distributions are shown in SI Fig. S3. They are largely consistent with the steady state results, broadening with decreasing $f_R$. They also show a small population of highly stretched rings at $\Delta x / L \approx 0.55 - 0.65$ for $f_R = 0.17,0.02$ corresponding to the extension overshoots. In the remainder of this work, we seek to understand the dynamics of ring polymers in blends with linear chains on  the basis of transient intermolecular interactions.

\section{\label{sec:Discussion}Discussion}

We consider two mechanisms for ring conformational fluctuations: i) intermolecular topological constraints in which a ring polymer is `hooked' or `threaded' by a linear polymer or another ring ii) intermolecular hydrodynamic interactions which fluctuate in space and time with the local concentration of linear and ring polymers. As discussed in the introduction, the first mechanism is well motivated by the observation of unique dynamics and rheology in ring polymer solutions and melts from bulk rheology, \cite{kapnistos2008unexpected,huang2019unexpected} single molecule experiments, \cite{robertson2007self,chapman2012complex} and molecular simulation. \cite{tsalikis2014threading,oconnor2020topological} A focus of this work is to determine if flow increases the frequency and importance of ring-linear threading such that the ring dynamics are significantly altered, despite the fact that  equilibrium ring dynamics at $c^*$ appear to be unaffected by threading. \cite{chapman2012complex}.  To understand the effect of HI, we visualize the flow as a function of time and make direct connections to fluctuations in ring polymer conformations. We find that flow modification is essential to understanding the polymer dynamics. 

\subsection{\label{sec:hooking}Intermolecular hooking}

We observe three types of topological interactions: ring-linear hooks, ring-ring hooks, and linear-linear hooks. We find ring-linear hooks to be the most common. Linear-linear hooks are less common and thus require a larger ensemble to quantify. We use data from simulation of pure linear polymer solutions at $1c^*$ and $3c^*$ previously published by the authors \cite{young2019simulation} while forgoing the blend cases due to insufficient sampling. Ring-ring hooks are the least common, such that we are unable to quantify their frequency. We include an observation of ring-ring hooking and a short discussion in SI Fig. S4.

\subsubsection{Observation of a ring-linear hook}

\begin{figure*}[htb]
	\includegraphics[width=\textwidth]{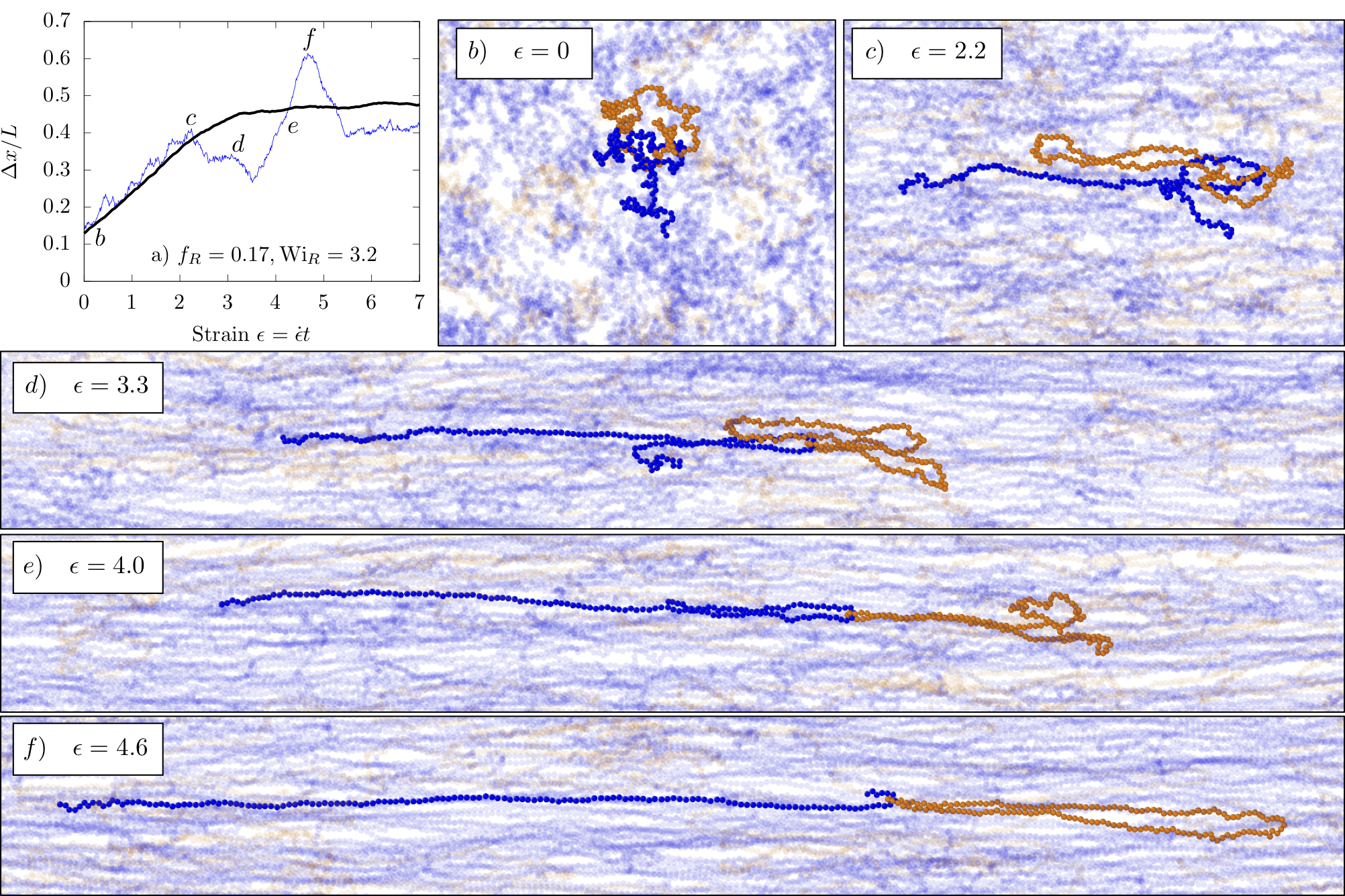}
    \caption{a) Transient fractional extension for a ring linear polymer which is threaded with a linear chain b) Initially, the linear chain is threaded `around' the outside of the ring c) For low accumulated strain up to $\epsilon \approx 2$, the ring does not feel the constraint and stretches affinely d) the hook inhibits stretching and causes the ring to retract and reorient e) The hook further stretches the ring in until it overshoots the ensemble average and eventually f) the hook is released and the ring returns to the average level of extension.
    }
    \label{fig:rlhook}
\end{figure*}

We first show simulation snapshots of a ring-linear hook to provide a qualitative description of the conformational dynamics. For examples of linear-linear hooks, we refer to our previous work. \cite{young2019simulation} A linear polymer is loosely threaded through a ring polymer upon the startup of flow (Fig \ref{fig:rlhook}b). At equilibrium, the excluded volume force between the ring and linear polymer is weak due to the low concentration. At low strain $\epsilon = 2.2$, the polymers have not yet collided so the ring stretches at approximately the same rate as the ensemble average (Fig \ref{fig:rlhook}c). Upon further strain accumulation, however, the crossing constraint sets in, and the linear chain adopts a folded conformation around the outside loop of the ring, causing it to retract and reorient (Fig \ref{fig:rlhook}d). The ring then becomes fully reoriented and overshoots the ensemble average fractional extension before the hook is released and the ring retracts to average levels of extension (Fig \ref{fig:rlhook}e,f).

For demonstrative purposes, we have highlighted a trajectory in which the ring both retracts and stretches due to the topological constraint. However, this is a rare occurrence. The majority of ring-linear hooks involve a linear polymer threaded `though' the ring (Supplementary Information movie I)  rather than `around', in which case the initial retraction does not occur. Additionally, in this example the linear polymer is already threaded with the ring upon startup of flow. This is the case in the majority of hooks we observe, although it is not required. In Supplementary Information movie I we include a trajectory in which an initially unthreaded linear chain adopts a folded conformation, advects into the closed contour of the ring, forms a topological hook that causes an overshoot in ring extension, and then advects away from the ring once the linear chain becomes fully stretched and the constraint is released. 

We note that rings tend to hook only with linear chains that adopt `folded' or `dumbbell' conformations. \cite{perkins1997single} Linear chains adopting `kinked' or `coiled' conformations lack a hooked structure at their free ends which may advect into the ring. These latter conformations can still thread through the ring, but in this case the excluded volume force is weak because the low polymer concentration allows the ring to deform and relax the constraint. Thus, there appears to be aspects of `molecular individualism' as first considered for dilute solution polymer stretching which are relevant in the semidilute case. Due to limited sampling and high computational expense, we are unable to study these problems in further detail here, although there may be similar instances of `molecular predestination'. \cite{larson1999brownian}

\subsubsection{Quantifying hooking behavior}

We now establish a procedure for detecting hooks and quantifying how often they occur. For the ring-linear and linear-linear hooks, we consider a combination of topological strand crossings and excluded volume interactions between hooked chains. In particular, we use the writhe for two polymer chains $\alpha$ and $\beta$
\begin{equation}
    \begin{split}
    Wr_{\alpha \beta} & = \frac{1}{4 \pi} \int_{C_{\alpha}} \int_{C_{\beta}} d \bm{r}_1 \times d \bm{r}_2 \cdot \frac{\bm{r}_1 - \bm{r}_2}{\left | \bm{r}_1 - \bm{r}_2 \right |^3} \\
    & \approx \frac{1}{4 \pi} \sum_{i=1}^{N_{\alpha}} \sum_{j=1}^{N_{\beta}} \frac{\bm{r}_i - \bm{r}_j}{r_{ij}^3} \cdot (\bm{r}_i \times \bm{r}_j)
    \end{split}
\end{equation}
where the line integrals over the continuous polymer curves $C_{\alpha}$ and $C_{\beta}$ are approximated by the discrete bead positions. In the case of a ring, the curve is closed, whereas for a linear chain it is open. While the writhe gives a measure of topological crossings, we are primarily interested in cases where these threads drive changes in the polymer conformation through excluded volume interactions. Thus, we also consider the flow-direction total excluded volume force between the two chains
\begin{equation}
    F_{x,\alpha \beta}^{EV} = \sum_{i=1}^{N_{\alpha}} \sum_{j=1}^{N_{\beta}} F_{x,ij}^{EV}
\end{equation}
with the excluded volume force between segments given by Eq. \ref{EV}. We set a threshold writhe $| Wr_h | = 1$ and excluded volume force $| F_{x,h}^{EV}| = 10 kT/a$, which when both are met indicate the presence of a hook. We then determine the writhe and excluded volume force between all chains on a pairwise basis for the simulation trajectories presented in the previous section and assign them a hook status $h_{\alpha \beta}(t)$ as a function of time.

We note that it is challenging to detect ring-ring hooks in this procedure because one half of the penetrating ring will form a positive crossing and the other half a negative crossing, yielding a writhe of zero. Instead, we use only the excluded volume condition for ring-ring hooks and confirm by visual observation. We find ring-ring hooks are nearly negligible at $c^*$, in agreement with equilibrium melt simulations showing ring-ring threads are significantly less probable than ring-linear threads. \cite{tsalikis2016analysis}

The topological constraint analysis reveals that once linear chains are fully stretched, ring-linear hooks are negligible. We explain this by the fact that a linear chain must have a significant end retraction of at least ~1/3 its steady state extension to be able hook with a ring as in the startup flow case. As shown in Fig. \ref{fig:fluc_avg} and Fig. S1, conformational fluctuations of linear polymers for $\textrm{Wi}_L > 1$ are small due to the large flow gradient across their span. Because of the difference in ring and linear polymer relaxation times ($\tau_L \approx 3.5 \tau_R$), the strain rates applied to the blends correspond to $\textrm{Wi}_L = 1.4 - 11.0$. Therefore, we focus on the startup of flow $t=t_0, \epsilon = 0$ to $t=t_{ss}, \epsilon=8$. In particular, we consider the number of ring-linear hooks per ring polymer, defined as
\begin{equation}
    \langle n_h \rangle_{t,n} = \frac{1}{n_{run} n_R (t_{ss} - t_0)} \sum_{i=1}^{n_{run}} \sum_{\alpha=1}^{n_R(i)} \sum_{\beta=1}^{n_L(i)} \sum_{t_0}^{t_{ss}} h_{\alpha \beta}(i,t)
\end{equation}
where the average is taken over time for the ensemble of ring-linear pairs at a given blend fraction and strain rate. The first summation is taken over the number of independent simulation runs. This definition accounts for the possibility of multiply-hooked rings, although they are not found in our simulations.

\begin{figure}[tb]
	\includegraphics[width=0.5\textwidth]{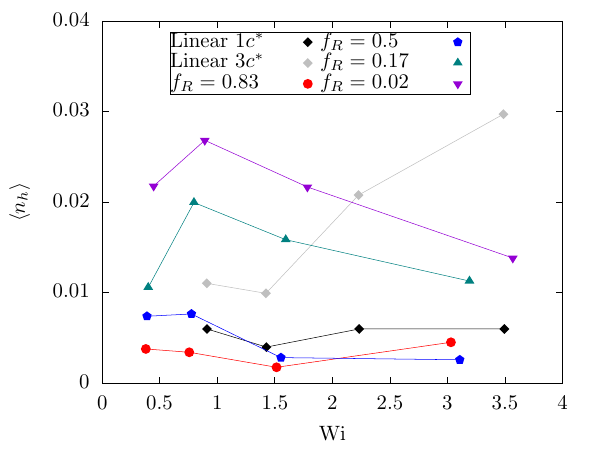}
    \caption{Time averaged density of ring-linear polymer intermolecular hooks as a function of Wi for varying blend ratio of rings $f_R$. The density of linear-linear polymer intermolecular hooks in a pure linear solution as determined from previous work by the authors\cite{young2019simulation} is included for comparison. The Weissenberg number Wi is defined with the appropriate relaxation time for either a ring or linear polymer at the relevant concentration and blend ratio.
    }
    \label{fig:hden}
\end{figure}

Fig. \ref{fig:hden} shows the results of the procedure as a function of flow rate and blend fraction. First we note the small quantitative values, reaching a maximum of $\langle n_h \rangle \approx 0.03$ for the case of a single ring in a linear background, indicating that at a given time during startup of flow, only 3\% of rings are hooked. Therefore, it appears that topological interactions alone cannot explain the large ring conformational fluctuations observed in simulation. Detailed quantitative study of the transient evolution of hook density and duration remain challenging, although several trends emerge.

A clear result is that the average number of hooks increases with blend fraction. This is expected, as rings form hooks with linear chains more readily than other rings. Majority ring polymer blends $f_R = 0.83,0.50$ exhibit nearly the same number of ring-linear hooks as linear-linear hooks in a pure linear polymer solution at the same concentration. For majority linear polymer blends $f_R = 0.17,0.02$, the average number of hooks increases by as much as a factor of 5 at low flow rates. We emphasize that $\langle n_h \rangle$ is determined on a per ring basis, such that blends at lower $f_R$ do not necessarily exhibit a larger number of ring-linear hooks per solution volume.

Another clear trend is the decreasing number of ring-linear hooks for solutions at $1c^*$ with flow rate.  This is in distinct contrast to the pure linear polymer solution at $3c^*$. We suggest that a crossover in the behavior of topological constraints in strong flows occurs in the range of $c/c^* \approx 1-3$. At $1c^*$, linear polymers threaded through a ring upon startup of flow tend to become fully stretched on a faster time scale than a hook forms for $\textrm{Wi}_R > 2$. Only initial configurations which are `predestined' to hook due to the threading of a folded linear polymer through a ring, for example in Fig. \ref{fig:rlhook}, form constraints. At lower flow rates, linear chains remain relatively coiled, allowing the constraint time to deform the ring. In the pure linear polymer solution at $3c^*$, the stretching of polymers to escape the constraint is limited by the surrounding chains. The entanglement concentration for the $N_L = 150$ linear polymer solutions is $c_e \approx$ 8-10 $c^*$, and the equilibrium diffusion follows unentangled scaling at $3c^*$. \cite{young2018conformationally}

We have not considered the magnitude of excluded volume force imposed by the constraint or the resulting increase in spring forces. Further studies with higher molecular weight polymers may reveal the functional effect of ring-linear hooks upon ring conformation and solution stress.

\subsection{\label{sec:himod}Flow modification by intermolecular hydrodynamic interactions}

 To explain the ring conformational fluctuations not caused by ring-linear hooking, we consider fluctuations in the effective flow due to the polymer disturbance velocity. We highlight several observations of rings exhibiting large fluctuations in extension and visualize the flow surrounding their centers of mass, defined as
\begin{equation}
    \bm{v}^*(\bm{r}_i) = \nabla \textbf{v} \cdot (\bm{r}_i - \bm{r}_{CoM}) + \sum_{j}^{'} \textbf{D}(\bm{r}_{ij}) \bm{F}_j
    \label{eqn:tot_flow}
\end{equation}
where $\bm{r}_i - \bm{r}_{CoM}$ is the displacement from a tagged ring center of mass $\bm{r}_{CoM}$, $\bm{r}_{ij} = \bm{r}_j - \bm{r}_i$, and $\bm{r}_j$ and $\bm{F}_j$ are the positions and total conservative force respectively of polymer bead $j$. The first term accounts for the applied planar extensional flow in the frame of reference of the ring center of mass. Because the applied flow is homogeneous and unbounded, we can define a new origin $\bm{r}_{CoM}$ to provide a more intuitive flow field without qualitatively changing the flow measurement. The second term accounts for the polymer disturbance velocity where the sum is over polymer bead indices, and the prime indicates that beads on the tagged ring are excluded. Intramolecular HI, or the ring's response to the flow, are excluded in order to visualize the forces which drive ring fluctuations.

 We observe several characteristic motions we refer to as i) overstretching ii) retraction iii) tank-treading. The first two cases refer to fluctuations in fractional extension more than the three times average fluctuation quantity quantified in Section \ref{sec:avg_fluc}, $ | \Delta x(t)/L - \langle \Delta x/L \rangle | > 3 \langle \delta \rangle$, either above or below the steady state average respectively. The third motion refers to rotation of the ring along its contour in the flow-neutral $z$-direction. 

We evaluate $\bm{v}^*$ on a uniform rectilinear mesh grid at locations $\bm{r}_i$ surrounding the ring center of mass. We find that snapshots of the flow field on the time scale of a single time step $dt = 5 \times 10^{-4} \tau_0$ are noisy due to thermal fluctuations. To overcome this, we perform additional simulations which sample the disturbance velocity every 50 time steps for an interval $50 \tau_0$ or $0.17 \tau_R$. Polymer advection is minimal on this time scale and the resulting streamlines are a close approximation to the instantaneous flow. The flow sampling increases the computational expense of the simulation by $A n_p N$ where $A$ is a constant associated with evaluating the RPY tensor for a grid point polymer bead pair, and $n_p$ is the number of grid points. As $n_p$ exceeds $N$ the simulation becomes computationally expensive, so we consider a volume extending only slightly past the extents of the tagged ring. Specifically, we consider the dimensions of the grid $\tilde{\bm{L}}_g = (120.0,40.0,48.0)$, and larger 2D slices in the flow-compression plane in Fig. \ref{fig:flowmod}. The average size of the simulation cell over a deformation period is $\tilde{\bm{L}} = (483.3,184.7,100.0)$. For visualization purposes we show only 2D slices of the flow and assume the flow in the $z$ direction is zero. This is generally a reasonable choice in the $xy$-plane, as the applied flow is planar, although significant gradients in the flow-neutral $z$-direction are possible (Fig. \ref{fig:hi_tt}).  Visualization of the full 3D flow field is an interesting area for further study but is unnecessary for understanding the dynamics of individual ring polymers.

The effective velocity is normalized by a reference value $U = \textrm{max}(| \nabla \textbf{v} \cdot (\bm{r}_i - \bm{r}_{CoM})|)$ corresponding to the maximum applied flow magnitude on the mesh. For positions further from the ring the applied flow becomes stronger relative to the polymer disturbance, but these flows are not relevant to the ring conformation. We now consider several specific examples of the characteristic motions described above and visualize the surrounding flow fields.

\subsubsection{Overstretching}

\begin{figure*}[htb]
	\includegraphics[width=\textwidth]{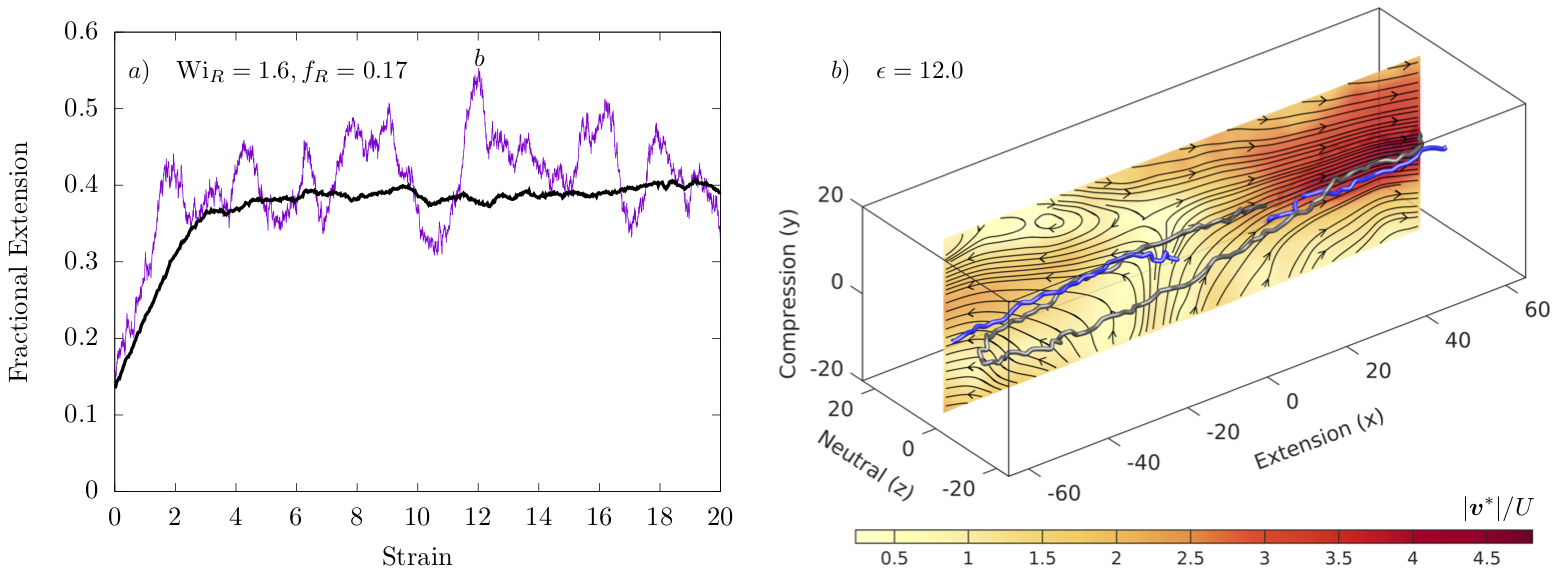}
    \caption{Ring polymer overstretching a) Transient fractional extension of the tagged ring b) Ring (grey) and linear (blue) polymer conformations superimposed on a slice of the normalized effective flow streamlines $\bm{v}^* / U$ in the $xy$-plane at a location $\tilde{z}=-4$ with respect to the ring center of mass. The snapshot is taken at $\epsilon = 12.0$ corresponding to the maximum in ring polymer stretch. The disturbance velocity of the linear polymers exceeds the underlying applied flow and causes the ring to stretch beyond the steady state average. Other neighboring polymers have been excluded for visual clarity. The ring is slightly stretched in the $z$-direction such that the strong extension at $\tilde{x} \approx -50$ is not visible in this slice, but the flow is similar to the opposite `end' of the ring at $\tilde{x} \approx 50$. The flow rate is $\textrm{Wi}_R = 1.6$ and the blend fraction is $f_R = 0.83$.
    }
    \label{fig:hi_hs}
\end{figure*}

In Fig. \ref{fig:hi_hs}a we show the transient fractional extension of a ring in a majority linear blend $f_R = 0.17$ at $\textrm{Wi}_R = 1.6$ which fluctuates around the ensemble average extension. At $\epsilon \approx 12.0$, the ring becomes highly stretched to $\Delta x(t)/L - \langle \Delta x / L \rangle > 3 \delta$. We plot the streamlines for the effective velocity in the $xy$-plane at this time superimposed with the ring conformation and selected linear conformations in Fig \ref{fig:hi_hs}b. The effective flow at the ring end $\tilde{x} \approx 50$ is $\sim$ 4.5 times stronger than the applied flow alone, causing the ring to stretch far beyond its average extension. We see that this clearly coincides with the position of a linear chain end. At the opposite end of the ring, $\tilde{x} \approx -50$ the ring is similarly stretched by a linear chain end, although this is not visible in the displayed slice because the ring is slightly stretched in the $z$-direction.

Thus, we find that local fluctuations in the polymer concentration drive flow modifications and determine ring conformational dynamics. In particular, the proximity of linear chain ends to ring `ends' can cause overstretching. While the highlighted linear chain ends do not completely reproduce the total flow, they contribute the essential features of strong extension. We note that only a section of the linear chains are shown. The full conformations extend in the flow direction away from the ring. If we consider the disturbance due to an individual polymer, the importance of the linear chain ends becomes clear. The restoring force on chain segments is strongest at the ends, while the disturbance at the chain center goes to zero.

\subsubsection{Retraction and tumbling}

\begin{figure*}[htb]
	\includegraphics[width=\textwidth]{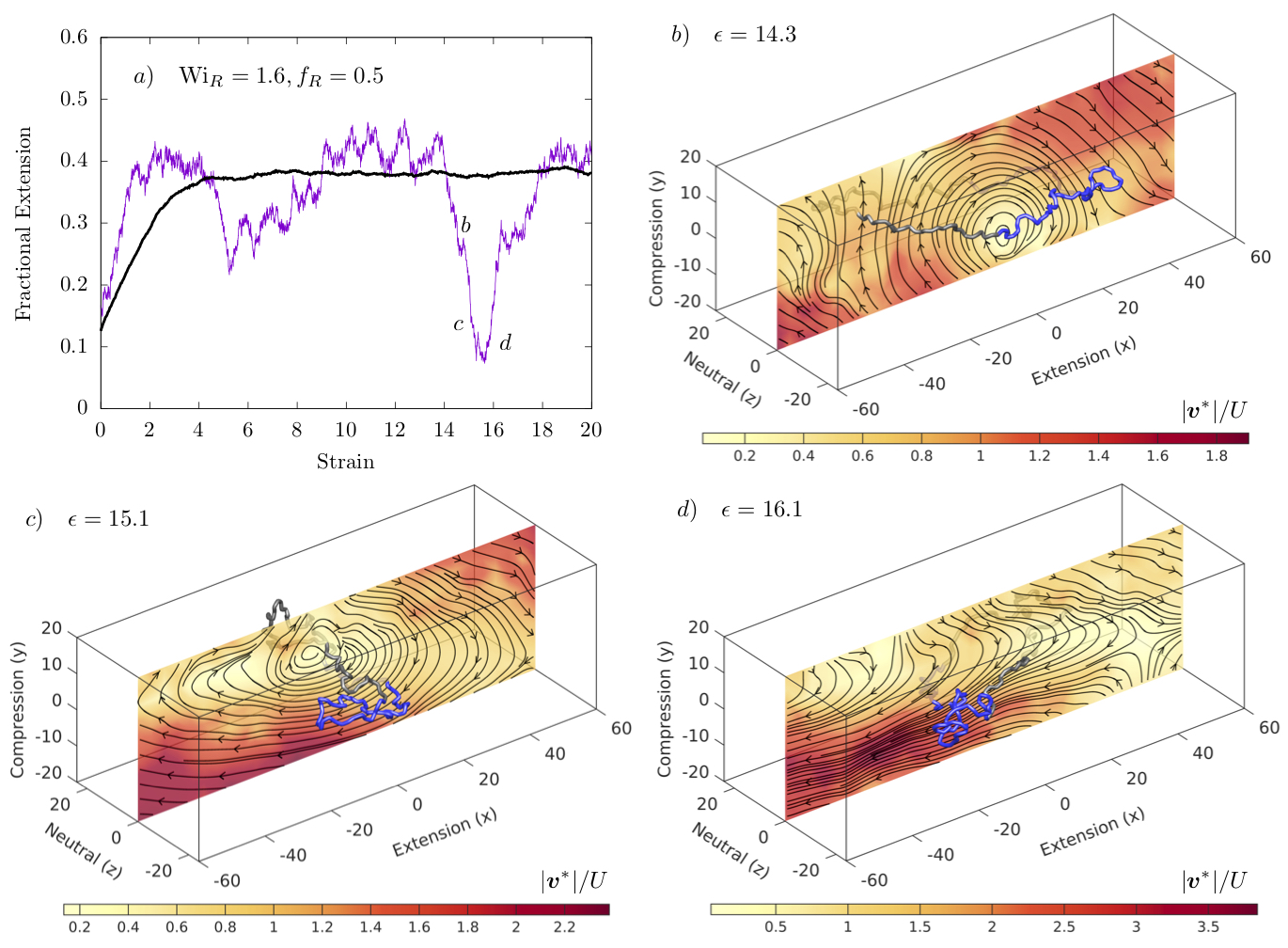}
    \caption{Ring polymer retraction and tumbling a) Transient fractional extension of the tagged ring b) Ring polymer conformation superimposed on a streamline slice in the $xy$-plane at $\tilde{z} = 0$ with respect to the ring center of mass. Half of the ring is colored blue and the other half grey to show tumbling. The ring is initially stretched in the flow direction before entering the rotational field c) The flow causes the ring to retract and reorient while swelling slightly in the $y$ direction so the `ends' can flow past each other d) The flow returns to approximately planar extension and the ring stretches back to average levels extension after complete reorientation. The flow rate is $\textrm{Wi}_R = 1.6$ and the blend fraction is $f_R = 0.5$.
    }
    \label{fig:hi_tumble}
\end{figure*}

Next we consider an example of ring retraction at $f_R = 0.5, \textrm{Wi}_R = 1.6$. The ring is initially stretched in the flow direction at $\epsilon \approx 13$, at which point a rotational flow field emerges (Fig \ref{fig:hi_tumble}b). The flow causes the ring to retract and tumble, passing through a minimum in fractional extension at $\epsilon \approx 15.3$ (Fig \ref{fig:hi_tumble}c). Finally, the flow surrounding the ring returns to approximately planar extension with a stagnation point laterally displaced to $\tilde{x} \approx 40$ (Fig \ref{fig:hi_tumble}d). The ring then completes reorientation in the flow direction and stretches back to average levels of extension. To facilitate visualization of this process, we have colored one half of the ring conformation grey and the other blue, showing that from $\epsilon = 14.3$ to $\epsilon = 16.1$ the two halves essentially exchange places.

Not all cases of ring retraction involve end-over-end tumbling. More often, the ring extension decreases by only  $\Delta x(t)/L - \langle \Delta x / L \rangle \approx 1-2 \langle \delta \rangle$ before restretching. This is due to the degree of flow modification, with weaker retractions corresponding to a mixed flow between rotation and extension, and tumbling corresponding to rotational flow.

\begin{figure*}[htb]
	\includegraphics[width=\textwidth]{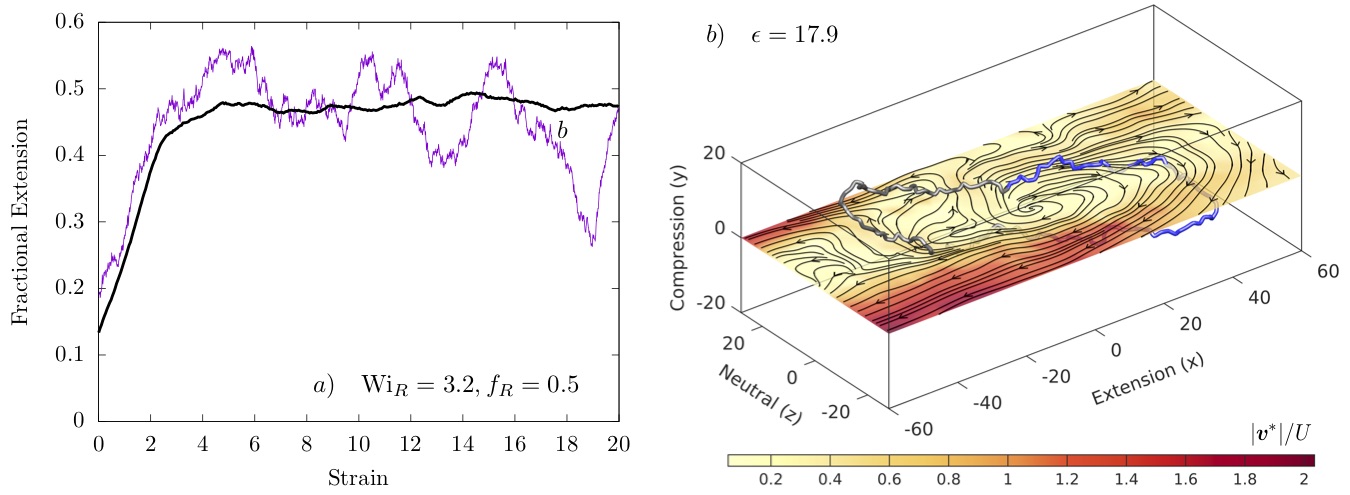}
    \caption{Ring polymer tank-treading a) Transient fractional extension showing a sharp retraction b) Ring polymer conformation superimposed on a streamline slice in the $xz$-plane at $\tilde{y} = 0$ with respect to the ring center of mass. The ring is initially stretched in an open loop conformation. The flow in the neutral $z$-direction is rotational, while in the $xy$-plane it is primarily extensional (not shown), causing the ring to rotate along its contour while remaining stretched in the $x$-direction. Upon further strain accumulation, the flow returns to planar extension as the neighboring chains advect. The flow rate is $\textrm{Wi}_R = 3.2$ and the blend fraction is $f_R = 0.5$.
    }
    \label{fig:hi_tt}
\end{figure*}

Generally, shear and rotational flows appear to emerge from the superposition of polymer disturbance velocities. In dilute solution, polymers align along the axis of extension, introducing only symmetric disturbance flows around the hydrodynamic center of resistance. However, these disturbance flows are not necessarily symmetric with respect to the stagnation point. Superposition of the disturbance from multiple chains shows significant deviations from pure extension. Thus, polymers can become unaligned with the axis of extension due to intermolecular HI. This observation is supported by experiments \cite{haward2016elastic} and numerical simulations \cite{cruz2018characterization}, which report that birefringence decreases with polymer concentration and molecular weight at a fixed Wi. While the origin of this behavior is clear from a continuum perspective,  \cite{shaqfeh1996purely, pakdel1996elastic} the molecular mechanism is not fully resolved. Further study is required to investigate the importance of intermolecular HI, coupled thermal fluctuations, and EV.

\subsubsection{Neutral direction gradients and tank-treading}

In both the overstretching and tumbling cases, we see that the ring is slightly stretched in the flow-neutral $z$-direction. This is not unexpected, as previous simulations have shown that the coupling of chain architecture and intramolecular HI in rings drives the chain to an open loop conformation. \cite{hsiao2016ring} The stretching of the ring is the cause of the mild coil-stretch transition and delay in critical Weissenberg number as compared to linear polymers. In the semidilute case, this may be complicated by the screening of HI, which causes rings to compress in the neutral direction as seen in dilute freely draining simulations.  \cite{hsiao2016ring} Alternatively, the concentration gradients which lead to the mixed flows seen in the extension-compression plane in Figs. \ref{fig:hi_hs} and \ref{fig:hi_tumble} could introduce flow gradients in the neutral direction which are absent in the dilute case.

We find that flow gradients in the neutral direction lead to ring retraction even when the flow the $xy$-plane remains primarily planar extensional. In Fig. \ref{fig:hi_tt}, we highlight a case where the ring retracts and partially `tank-treads', meaning the ring rotates along its contour while remaining stretched in the flow direction. We include a simulation movie in the Supplemental Information. Tank-treading has been reported for dilute rings in shear flow, where the driving force for rotation is the shear gradient on ring polymers adopting elliptical conformations in the flow-gradient plane. \cite{chen2013tumbling} In this work, the dynamics are markedly different. The applied flow is planar extensional, such that there is no component of rotation and the ring is compressed in the $y$ direction. The tank-treading motion can emerge only due to a combination of intramolecular HI driving the ring open in the neutral direction and intermolecular HI introducing a driving force for rotation.

The origin of rotation is clearly seen in Fig. \ref{fig:hi_tt}b, where the polymer disturbance velocity causes rotational flow in the $xz$ (extension-neutral) plane. At a later time, where the ring has rotated around its contour by $\approx$ \ang{90}. The flow then returns to planar extensional. This evolution of the flow field is directly connected to the motion of neighboring chains. In the simulation movie we highlight three linear polymers which primarily drive the flow. At $\epsilon = 18.1$ (start of the movie), two linear chains appear overlapping in the extension-compression plane but with a separation in the neutral direction of approximately the ring stretch, $\Delta z_R \approx z_{L1} - z_{L2}$. The linear chains are highly stretched with nearly constant $z$ positions along their contours. The two linear chains are advecting in opposite directions with respect to the axis of extension due to their center of mass positions in the applied flow. The superposition of their disturbance velocities leads to the rotational field. Once the linear chains have advected away from the ring, their disturbance decays and the rotational field dissipates. 

More generally, the role of flow gradients in the $z$ direction is challenging to characterize. Tank-treading is rare because the spatiotemporal variations of the flow occur on time scales shorter than the time it takes the ring to rotate along its contour. We conclude that the conformational fluctuations as measured in a coarse-grained sense by the average fluctuations in the fractional extension $\langle \delta \rangle$ are due to diverse features of the flow which arise from the polymer disturbance velocity. We now discuss some more general features of deviations from planar extensional flow and relate them to the dynamics of ring polymers.


\subsubsection{\label{sec:bulkflow}Deviations from planar extensional flow}

\begin{figure}[htb]
	\includegraphics[width=0.45\textwidth]{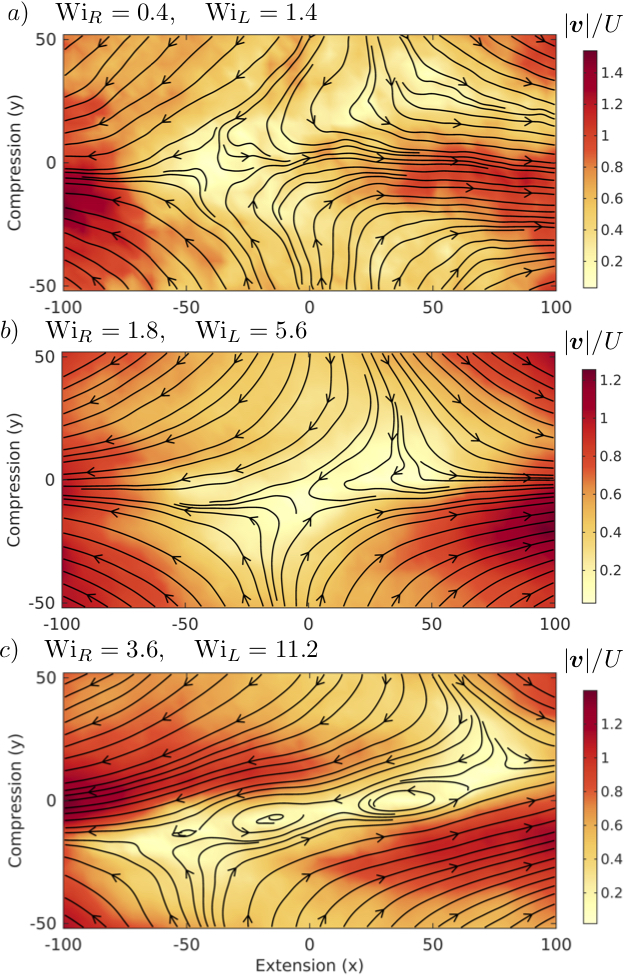}
    \caption{Total effective flow due to applied PEF and polymer disturbance for a trace ring in a linear semidilute background $f_R = 0.02$ at flow rates a) $\textrm{Wi}_R = 0.5, \textrm{Wi}_L = 1.4$ stagnation point displacement b) $\textrm{Wi}_R = 1.8, \textrm{Wi}_L = 5.6$ transient flow asymmetry c) $\textrm{Wi}_R = 3.6, \textrm{Wi}_L = 11.2$ transient flow asymmetry. Streamlines are averaged for 2 units of Hencky strain.
     }
    \label{fig:flowmod}
\end{figure}

The observation of rotational flows in Fig. \ref{fig:hi_tumble} is a marked change in the flow character as compared to the applied planar extension. To further investigate these profiles, we consider larger slices surrounding the stagnation point in Fig. \ref{fig:flowmod}. The method of measuring the flow is the same as Eqn. \ref{eqn:tot_flow}, except that the stagnation point is the reference rather than a tagged ring center of mass, and the flow disturbance from all polymers is included. To simplify the discussion, we consider only the results for nearly pure linear polymer solutions ($f_R = 0.02$), where the effect of chain architecture on the solution average behavior is small. Because there are few rings, we use the dimensionless flow strength based on the linear polymer relaxation time, $\textrm{Wi}_L$.

We observe several characteristic deviations from planar extension. At moderate flow rates $\textrm{Wi}_L \approx 1.4$, the stagnation point is displaced from that of the underlying flow. The displacement is primarily in the direction of extension, and smaller displacements are observed in the compression direction. In faster flows $\textrm{Wi}_L \geq 2.8$, symmetry across the axis of extension is broken and regions of shear flow emerge. The flow modification is most noticeable near the stagnation point, where the polymer disturbance velocity is large relative to the applied flow. Finally, regions of rotational flow are also observed (Fig. \ref{fig:flowmod}c). Vortices occur within regions of shear flow, and are generally smaller than the shear regions.

These are transient features which are not sustained for more than $\epsilon_H \approx 2$ units of accumulated strain. Once polymers fully stretch, the flow is rarely purely planar extensional, but is rather fluctuating between states similar to those shown in Fig. \ref{fig:flowmod}, as shown in Fig. S5. There is no preferred direction of stagnation point displacement, flow asymmetry, or vorticity. In particular, both clockwise (Fig. \ref{fig:hi_tumble}) and counter-clockwise (Fig. \ref{fig:flowmod}c) rotational flows are observed. While these deviations from planar extension are smaller than the periodic simulation cell size $\tilde{\bm{L}} = (483.3,184.7,100.0)$, finite size effects may be present because polymers can interact with their own images via long-ranged HI. Therefore, we defer a quantitative characterization of flow modification to a future study which uses more coarse-grained models to more fully investigate and quantify these transient flows.

Deviations from extensional flow are perhaps not surprising considering that elastic instabilities have been reported in cross-slot extensional flows of polymer solutions, \cite{arratia2006elastic,poole2007purely} although the flow asymmetry is stable and extends to the boundaries of the microfluidic device. Transient stagnation point displacement and rotational flows have also been reported in an optimized-shape cross-slot extensional rheometer, which more closely resemble homogeneous extensional flow. \cite{haward2016elastic,cruz2018characterization} However, the flow kinematics in a cross-slot geometry are different than in the BD simulations presented in this work. In cross-slot devices, only the flow in the cross-channel region is homogeneous extension, leading to inhomogeneous polymer stretching as indicated by a narrow birefringent strand along the center line. \cite{haward2016elastic,cruz2018characterization} The applied flow field in the current BD simulations is unbounded and homogeneous, and polymer stretching is relatively constant throughout the simulation cell. This suggests the shear and rotational deviations observed here may be generic to extensional flows of polymer solutions regardless of the boundary conditions.

\subsection{Test case: a bidisperse linear blend}

The observation that ring polymers exhibit large conformational fluctuations due to size and shape differences with the linear portion of the blend raises the question: what is the influence of chain architecture? Is there a unique feature of the ring polymers that causes these dynamics, or is it simply because their contour length is half that of the linear chain? To investigate this topic, we have performed simulations of bidisperse linear polymer solution blends of short $N_S = 75$ and long $N_L = 150$ linear polymers. The maximum stretch of the short chains in the extension direction is matched to the rings, $L_S = L_R = 0.5L_L$, and the relaxation time is comparable ($\tau_R = 274 \tau_0$ vs $\tau_S = 322 \tau_0$ in pure solution $f_R = 1$ and $f_S = 1$ respectively at $c_L^*$). The polymer concentration by mass $c = N/V$ is matched to the ring-linear blend simulations. We define the mass fraction of short linear chains $f_S = n_S N_S / (n_S N_S + n_L N_L)$, and simulations are performed at $f_S = 1,0.5$. Essentially, each ring polymer is replaced with two short linear chains, all chains are initialized as non-overlapping random walks, and the TEA parameters are modified appropriately. Otherwise, details of the simulation are the same as described in Section \ref{sec:Simulation Method}. We define a Weissenberg for the short linear polymers $\textrm{Wi}_S = \dot{\epsilon} \tau_S$, where $\tau_S$ is the short polymer relaxation time at the appropriate blend ratio and is determined by the same procedure as for the ring-linear blends.

\begin{figure}[tbh]
	\includegraphics[width=0.4\textwidth]{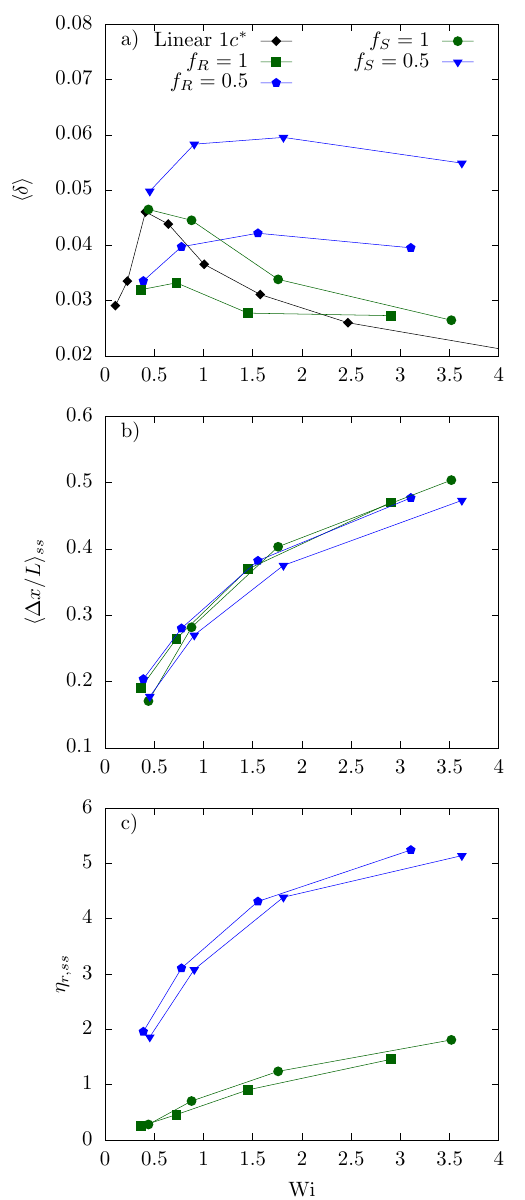}
    \caption{Ensemble average steady state a) fluctuation quantity $\langle \delta \rangle$ b) fractional extension $ \langle \Delta x / L \rangle_{ss}$ c) reduced extensional viscosity $\eta_{r,ss}$ as a function of Wi for ring-linear blends at $f_R = 1, 0.5$ as compared to bidisperse linear blends at $f_S = 1, 0.5$. The Weissenberg number is defined as the strain rate normalized by the corresponding relaxation time for the architecture and blend composition.
    }
    \label{fig:bidisp}
\end{figure}

In Fig. \ref{fig:bidisp} we compare the results of ring-linear blends and bidisperse linear blends at varying blend fraction and flow rate Wi, where the relaxation time $\tau_R$ or $\tau_S$ at the relevant blend fraction is used. The two systems are in qualitative and in some cases quantitative agreement. The primary result is that the fluctuation quantity for the short linear chains follows the trend of the rings. In the pure short linear polymer solution $f_S = 1$, $\langle \delta \rangle$ is peaked at $\textrm{Wi} = 0.5$ and monotonically decreases with Wi, in agreement with the reference data. \cite{young2019simulation} In the bidisperse linear blend $f_S = 0.5$, however, the fluctuation quantity matches the pure solution case at $\textrm{Wi} \approx 0.5$, increases up to $\textrm{Wi} \approx 1.5$, and decreases for stronger flows. We also find that the fluctuations are quantitatively larger than the ring-linear blend case. A clear explanation for this behavior is that at low Wi, the dynamics are weakly dependent on blend ratio. By nature ring polymer conformational fluctuations are smaller due to the lack of free ends, so $\langle \delta \rangle$ for rings simply starts at a lower value. The end-free constraint is relevant in strong flows as well. For example, short linear polymers in the bidisperse blend can undergo end-over-end tumbling due to flow gradients in the $z$-direction while remaining significantly compressed in the $y$-direction (see Supplementary Information movie IV for an example at $f_S = 0.5, \textrm{Wi}_S = 3.5$). This conformational pathway is unavailable to rings, however, which retract weakly or rotate along their contours as seen in Fig. \ref{fig:hi_tt}, but cannot tumble while fully compressed in the $y$-direction. Instead, rings must swell in the $y$-direction to completely tumble (Fig \ref{fig:hi_tumble}), a pathway that is suppressed by the compressional component of the applied planar extension. 

Pure linear polymer solutions at matched contour length with the rings ($N_S = 75, N_R = 150$) are stretched to the same extension as rings at the same Wi. However, the bidisperse linear blend $f_S = 0.5$ is less stretched. Interestingly, a similar result is obtained in experiment for ring-linear blends at $f_R = 0.83$ vs. $f_R = 0.5,0.17$. We explain this result by the larger fluctuation quantity $\langle \delta \rangle$ for the bidisperse linear blends, which are close to the quantitative values for rings observed in experiment. At high Wi, chains cannot stretch far beyond their average extension due to finite extensibility. Therefore, conformational fluctuations at high Wi correspond to retractions. For sufficiently large $\langle \delta \rangle$, this results in a lower $\langle \Delta x / L \rangle_{ss}$.

The extensional viscosity at both blend ratios matches nearly quantitatively for ring-linear and bidisperse linear blends. This supports our previous conclusion that the stress arises from the polymer stretch. Intermolecular hooking is virtually absent in the bidisperse linear blends $f_S = 0.5$, so in both blends the increase in $\eta_r$ with increasing fraction of $N_L = 150$ linear polymers is due to the higher stretch.

Overall, the bidisperse linear blend results show that the large conformational fluctuations we observe for the ring-linear case may be a more general feature in polymer solution blends and polydisperse  solutions. Polydisperse solutions are often modeled with two-state models, where polymers are assumed to be either stretched or coiled based on their relaxation time, and intermolecular interactions are neglected. Our results show that intermolecular interactions in semidilute solutions under strong flows are not screened. In fact, they can lead to qualitatively different dynamics. The introduction of non-linear polymer architectures further complicates the issue. While we observe topological constraints to be minimal in simulation, they likely play a more important role at higher concentrations and higher molecular weight, as suggested by the larger $\langle \delta \rangle$ in experiment. Even in the absence of topological constraints, the effect of architecture is not trivial. Polymer size and shape play an important role which is challenging to predict \textit{a priori} given the large concentration fluctuations present in semidilute solutions.


\section{\label{sec:Conclusions}Conclusions}

We have investigated the dynamics and rheology of ring-linear polymer solutions blends at the overlap concentration in planar extensional flow via Brownian dynamics simulations and single molecule experiments. Simulations and experiments both show that as the blend fraction of rings $f_R$ decreases from a pure ring solution, the conformational fluctuations of the ring polymer portion of the blend increase. Simulations reveal that the origin of these dynamics are a combination of intermolecular topological and hydrodynamic interactions. We show that application of strong flows $\textrm{Wi}_R > 1$ can cause strong topological constraints in which a linear chain threads through a ring and deforms it via an intermolecular `hook'. Equilibrium diffusion of rings at the overlap concentration is relatively insensitive to blend ratio, suggesting that flow introduces new dynamics. \cite{chapman2012complex} This is supported by the observation that pure linear solutions at $1c^*$ form significantly fewer intermolecular hooks than ring-linear blends at $f_R = 0.17,0.02$. Hooking leads to overshoots in transient fractional extension on startup of flow for individual ring trajectories, which is quantified by conformational distributions. However, in simulation we find that the average number of hooks per chain upon startup of flow is low ($n_h < 0.05$), and once linear chains stretch fully at $\epsilon \approx 8$, hooking is nearly negligible. Considering that experiments show larger fluctuations, this effect could be sensitive to molecular weight and details of the flow geometry.

We show that steady state conformational fluctuations in simulation are driven by intermolecular HI. Three characteristic ring motions are observed: overstretching, retraction, and tank-treading. We measure the effective flow field due to the applied planar extensional flow plus the polymer disturbance velocity to explain these dynamics. We find that fluctuations in local concentration modify the flow and introduce regions of shear, rotational, and enhanced extensional flows that cause ring deformation. We suggest that the flow modification is stronger for majority linear polymer blends because of the stronger restoring force, causing fluctuations to increase with decreasing blend fraction of rings. We directly test the influence of polymer architecture, size, and shape by comparing the ring-linear systems to bidisperse linear polymer blends in which ring polymers are replaced with linear chains of matched contour length. The dynamics of the short linear chains are in good qualitative and quantitative agreement with the rings. Thus, our simulations have broader relevance to polymer solution blends and polydisperse solutions in which intermolecular HI could drive unexpected dynamics.

The current work considered only solutions at $1 c^*$. The entanglement concentration for $\lambda$-DNA is $c_e \approx$ $3 c^*$, and for the simulation model we estimate $c_e \approx$ 8-10 $c^*$. Despite this low concentration, we see emergent flow-induced topological constraints which could lead to entanglement dynamics below the equilibrium $c_e$. A potential area for future study is to quantify the crossover to entangled dynamics in non-equilibrium solutions with non-linear architectures. More generally, it is of interest to qualitatively characterize non-equilibrium conformations and solution stress in ring-linear blends above $c^*$, as equilibrium single molecule experiments suggest unique dynamics emerge compared to pure ring or pure linear solutions. \cite{robertson2007self, chapman2012complex}

A more detailed understanding of the influence of intermolecular HI is also essential to elucidate the ring dynamics and bulk flow properties. While equilibrium scaling theories normally neglect intermolecular HI below $c^*$, there is considerable evidence from bulk rheology, \cite{clasen2006dilute, dinic2017pinch, dinic2020flexibility} molecular simulations, \cite{stoltz2006concentration, prabhakar2017effect, young2019simulation} and theory \cite{prabhakar2016influence} that as the pervaded volume of the polymer increases with strain rate, intermolecular interactions become relevant at concentrations significantly below $c^*$. Furthermore, a detailed study of flow-concentration coupling in connection to molecular dynamics is of interest. Here, we clearly show that spatiotemporal concentration fluctuations modify the effective flow and drive conformational dynamics. Flow modification could also be relevant to kinematically mixed flows, \cite{cromer2017concentration, corona2018probing} which should now be accessible by molecular simulation due to recent progress in algorithms for periodic boundary conditions.  \cite{hunt2010new,jain2015brownian}

\section*{Supplementary Material}

In the supplemental material, we include movies showing characteristic examples of:

\begin{enumerate}
	\item Ring-linear hooking at $\textrm{Wi}_R = 1.8, f_R = 0.02$
	\item Ring tank treading at $\textrm{Wi}_R = 3.2, f_R = 0.50$
	\item Short linear polymer tumbling at $\textrm{Wi}_S = 3.5, f_S = 0.50$
\end{enumerate}

\begin{acknowledgements}
This work was funded by the National Science Foundation under Grant No. CBET-1803757 for CES, a DuPont Science \& Engineering fellowship for CDY, the National Science Foundation (NSF) Award CBET-1604038 for CMS and partially supported by the NSF through the University of Illinois at Urbana-Champaign Materials Research Science and Engineering Center (MRSEC) DMR-1720633 (YZ and CMS). The authors thank Sarit Dutta for helpful discussions.
\end{acknowledgements}

\appendix

\bibliography{main}

\end{document}



\title[Supplementary Information: Dynamics and Rheology of Ring-Linear Blend Semidilute Solutions in Extensional Flow: Modeling and Molecular Simulations]
{Supplementary Information: Dynamics and Rheology of Ring-Linear Blend Semidilute Solutions in Extensional Flow: Modeling and Molecular Simulations}

\author{Charles D. Young}
\affiliation{Department of Chemical and Biomolecular Engineering, University of Illinois at Urbana-Champaign, Urbana, Illinois 61801, USA}
\affiliation{Beckman Institute for Advanced Science and Technology, University of Illinois at Urbana-Champaign, Urbana, IL, 61801}
\author{Yuecheng Zhou}
\altaffiliation{Current address: Department of Chemistry, Stanford University, Stanford, California 94305, USA}
\affiliation{Department of Materials Science and Engineering, University of Illinois at Urbana-Champaign, Urbana, Illinois 61801, USA}
\affiliation{Beckman Institute for Advanced Science and Technology, University of Illinois at Urbana-Champaign, Urbana, IL, 61801}
\author{Charles M. Schroeder}
\affiliation{Department of Materials Science and Engineering, University of Illinois at Urbana-Champaign, Urbana, Illinois 61801, USA}
\affiliation{Department of Chemical and Biomolecular Engineering, University of Illinois at Urbana-Champaign, Urbana, Illinois 61801, USA}
\affiliation{Beckman Institute for Advanced Science and Technology, University of Illinois at Urbana-Champaign, Urbana, IL, 61801}
\author{Charles E. Sing}
 \email{cesing@illinois.edu}
\affiliation{Department of Chemical and Biomolecular Engineering, University of Illinois at Urbana-Champaign, Urbana, Illinois 61801, USA}
\affiliation{Beckman Institute for Advanced Science and Technology, University of Illinois at Urbana-Champaign, Urbana, IL, 61801}

\date{\today}

\maketitle

\section{Iterative Conformational Averaging method for ring-linear polymer solution blends}

We follow the approach of Geyer and Winter, who introduced the truncated expansion ansatz (TEA) approximation to the correlated Brownian noise. \cite{geyer2009n} The CA method introduces two further assumptions: (i) the decomposition coefficients of the TEA are conformationally averaged (ii) for interparticle distances above a cutoff $\tilde{r}_c$, the RPY tensor is discretely evaluated on a grid. The Langvein equation then becomes
\begin{equation}
    \label{eqn:LangevinCA2}
    \frac{d\tilde{\bm{r}}_{i}^{(w)}}{d\tilde{t}} = \tilde{\bm{\kappa}} \cdot \tilde{\bm{r}}_{i}^{(w)} -\sum_{j} \tilde{\textbf{D}}_{ij}^{eff} \nabla_{\tilde{\bm{r}}_{j}}(\tilde{U}) +     \tilde{\bm{\xi}}_{i}^{(w)}(\epsilon_o)
\end{equation}
where the superscript $(w)$ denotes the iteration number. The diffusion tensor  is approximated by an exact Ewald sum within a cutoff radius $\tilde{r}_c = 12a$ and a discrete approximation to the RPY tensor outside the cutoff
\begin{equation}
    \tilde{\textbf{D}}_{ij}^{eff} = \tilde{\textbf{D}}_{ij}^{RPY} \Theta(\tilde{r}_{c} - \tilde{r}_{ij}) + \tilde{\textbf{D}}_{ij}^{G}(t) \Theta(\tilde{r}_{ij} - \tilde{r}_{c})
\end{equation}
The $RPY$ superscript indicates the full Ewald sum and the $G$ the discrete grid space approximation, given by $\tilde{\textbf{D}}_{ij}^{G} = \tilde{\textbf{D}}^{RPY}(\Delta\tilde{\bm{r}}_{ij})$, where $\Delta \tilde{\bm{r}}_{ij} = (\Delta \tilde{x}_{ij}, \Delta \tilde{y}_{ij}, \Delta \tilde{z}_{ij})$ is the pair displacement rounded to the nearest grid point. We use a grid spacing $d_g = 1.0a$ for all simulations. Note that the diffusion tensor does not include an iteration superscript $(w)$ because in this formulation the HI depends only on the inter-bead displacements and no preavaging is used. The diffusion tensor is updated every $\lambda^{RPY} = \Delta t n^{RPY} = 0.05 \tau_0$, where $n^{RPY} = 100$ is the number of time steps between updates.

The conformationally averaged form of the TEA gives the Brownian noise at an accumulated strain $\epsilon$ as
\begin{equation}
	\label{eqn:TEACA}
    \tilde{\xi}_{l}^{(w)}(\tilde{t},\epsilon) = \tilde{\textrm{D}}_{ll}^{eff}(\tilde{t}) \langle C_{l} \rangle ^{(w-1)} (\epsilon)  \langle \beta' \rangle ^{(w-1)} (\epsilon) \sum_{m=1}^{3N} \frac{ \tilde{\textrm{D}}_{lm} ^{eff}(\tilde{t})}{\tilde{\textrm{D}}_{ll}^{eff} (\tilde{t})} f_{m}(\tilde{t})
\end{equation}
Note the index $l$ gives an individual component of the size $3N$ noise vector and not three components $(x,y,z)$ for a bead $i$ as in Eqn. \ref{eqn:LangevinCA2}. The TEA is a pairwise approximation to the exact decomposition. The $\beta'$ parameter describes the average hydrodynamic coupling and the coefficients $C_l$ ensure the beads experience the correct Stokes drag. The average quantities sampled from the previous iteration are evaluated transiently to account for startup and relaxation transience, leading to
\begin{equation}
    \label{eqn:BTEA}
    \langle \beta' \rangle^{(w)}(\epsilon_{o}) = \frac{1}{T}\sum_{\epsilon_o t_{\epsilon_{bin}}}^{(\epsilon_o+1) t_{\epsilon_{bin}}}{\beta'^{(w)}(t)} \\
\end{equation}
\begin{equation}
    \label{eqn:CTEA}
    \langle C_{l} \rangle^{(w)}(\epsilon_{o}) = \frac{1}{T}\sum_{\epsilon_o t_{\epsilon_{bin}}}^{(\epsilon_o+1) t_{\epsilon_{bin}}}{C^{(w)}_{l}(t)}
\end{equation}
where $\epsilon_o$ refers to the strain bin, and the sums indicate sampling of a strain interval during the stretching phase and a time interval during the relaxation phase as detailed in the authors' previous work. \cite{young2019simulation} The instantaneous samples of the average are defined as
\begin{equation}
	\label{TEA3}
    \beta' = \frac{1-\sqrt{1-3N(\varepsilon^{2}-\varepsilon)}}{3N(\varepsilon^{2}-\varepsilon)}
\end{equation}
where $\varepsilon$ is an average over the off-diagonal entries of the diffusion tensor
\begin{equation}
	\label{TEA4}
    \varepsilon = \frac{1}{(3N)^{2}}\sum_{l} \sum_{m \ne l} \frac{\tilde{\textrm{D}}_{lm}}{\tilde{\textrm{D}}_{ll}}
\end{equation}
The coefficients are given by
\begin{equation}
	\label{TEA2}
    C_{l} = \sqrt{\frac{1}{1+\beta'^{2} \sum_{l} \sum_{m \ne l} \frac{\tilde{\textrm{D}}_{lm}^{2}}{\tilde{\textrm{D}}_{ll}\tilde{\textrm{D}}_{mm}}}}
\end{equation}
Further details are given by Geyer and Winter \cite{geyer2009n} and the authors' previous work. \cite{young2019simulation} We make minor modifications to the method for the blend case. In principle, there are 3 coefficients $C_l$ associated with each bead. Assuming polymers to be distinguishable only by their initial configurations, we previously used a ensemble averaged set of $3 N_L$ coefficients for each linear chain. For blends, there is an additional set of $3 N_R$ coefficients for rings. The $\beta'$ parameter is a solution averaged quantity that is not specific to the chain architecture. We perform two iterations to obtain freely draining (FD, $w=0$) and hydrodynamically interacting (HI, $w=1$) results. The authors have shown that the second iteration ($w=1$) provides excellent agreement with traditional BD simulations, and a third ($w=2$) iteration does not significantly improve accuracy. \cite{young2019simulation}

\section{Extension fluctuations of linear polymers}

In Fig. \ref{fig:lin_fluc}, we compare the conformational fluctuations of $N_L = 150$ linear polymers in a semidilute blend with ring polymers to a dilute and semidilute ($1c^*$) pure linear polymer solution. For $f_R \leq 0.83$, the results for a ring-linear blend nearly quantitatively match the pure $N_L = 100$ linear polymer solution. Semidilute solutions at all blend ratios exhibit larger conformational fluctuations than linear polymers in dilute solution due to intermolecular HI.

\begin{figure}[htb]
	\includegraphics[width=0.6\textwidth]{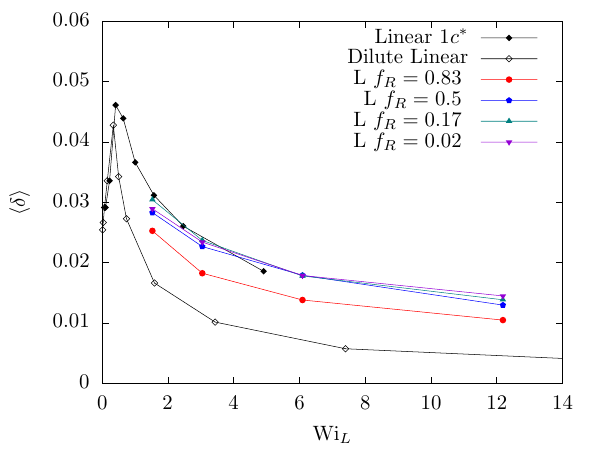}
    \caption{Conformational fluctuations of $N_L = 150$ linear polymers in a blend with ring polymers of varying blend ratio in comparison to $N_L = 150$ linear polymers in dilute solution and $N_L = 100$ linear polymers at the overlap concentration $c^*$.
    }
    \label{fig:lin_fluc}
\end{figure}

\section{Ring contribution to the extensional viscosity}

\begin{figure}[htb]
	\includegraphics[width=\textwidth]{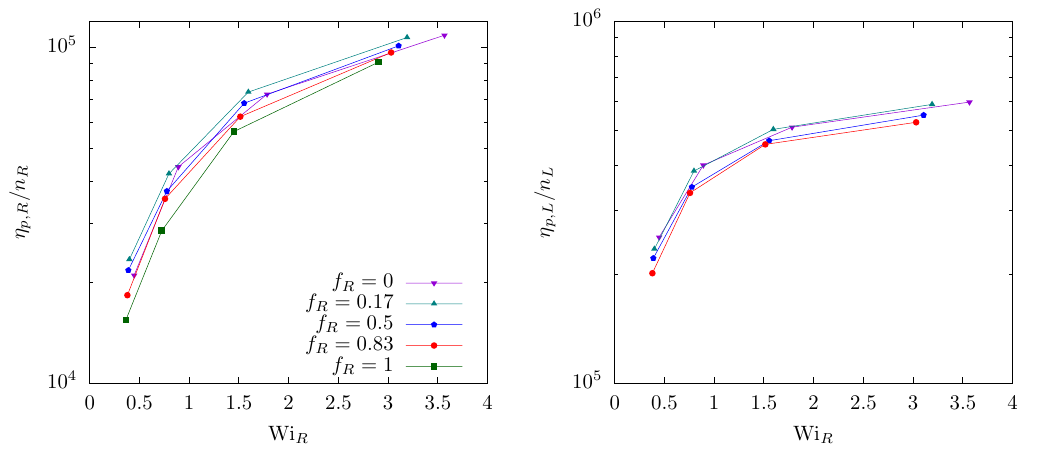}
    \caption{Contribution of a) ring and b) linear polymers to the extensional viscosity normalized by the number of ring and linear polymers respectively.
    }
    \label{fig:lin_fluc}
\end{figure}

The increase in extensional viscosity as  decreases is simply due to the higher stress from each linear chain. The contribution from each ring and linear chain does not significantly change with the blend ratio. We show this in Fig. 2, where we show the ring and linear contributions to the extensional viscosity:

\begin{equation}
	\eta_{p,R} = -\frac{\tau_{p,R,xx} - \tau_{p,R,yy}}{\dot{\epsilon}}
\end{equation}
\begin{equation}
	\eta_{p,L} = -\frac{\tau_{p,L,xx} - \tau_{p,L,yy}}{\dot{\epsilon}}
\end{equation}
where the ring and linear polymer stress tensors are
\begin{equation}
	\tau_{p,R,\alpha \beta} = \frac{1}{V} \sum_{i=1}^{n_R N_R} \sum_{j=1}^N r_{ij,\alpha} F_{ij,\beta}
\end{equation}
\begin{equation}
	\tau_{p,L,\alpha \beta} = \frac{1}{V} \sum_{i=1}^{n_L N_L} \sum_{j=1}^N r_{ij,\alpha} F_{ij,\beta}
\end{equation}
The first sum is over only ring polymer beads or only linear polymer beads respectively. When normalized by the number of molecules in solution, we find the ring and linear contributions to the extensional viscosity depend only weakly on blend ratio, with a slight increase as $f_R$ decreases. 

\section{Transient conformational distributions}

\begin{figure*}[htb]
	\includegraphics[width=\textwidth]{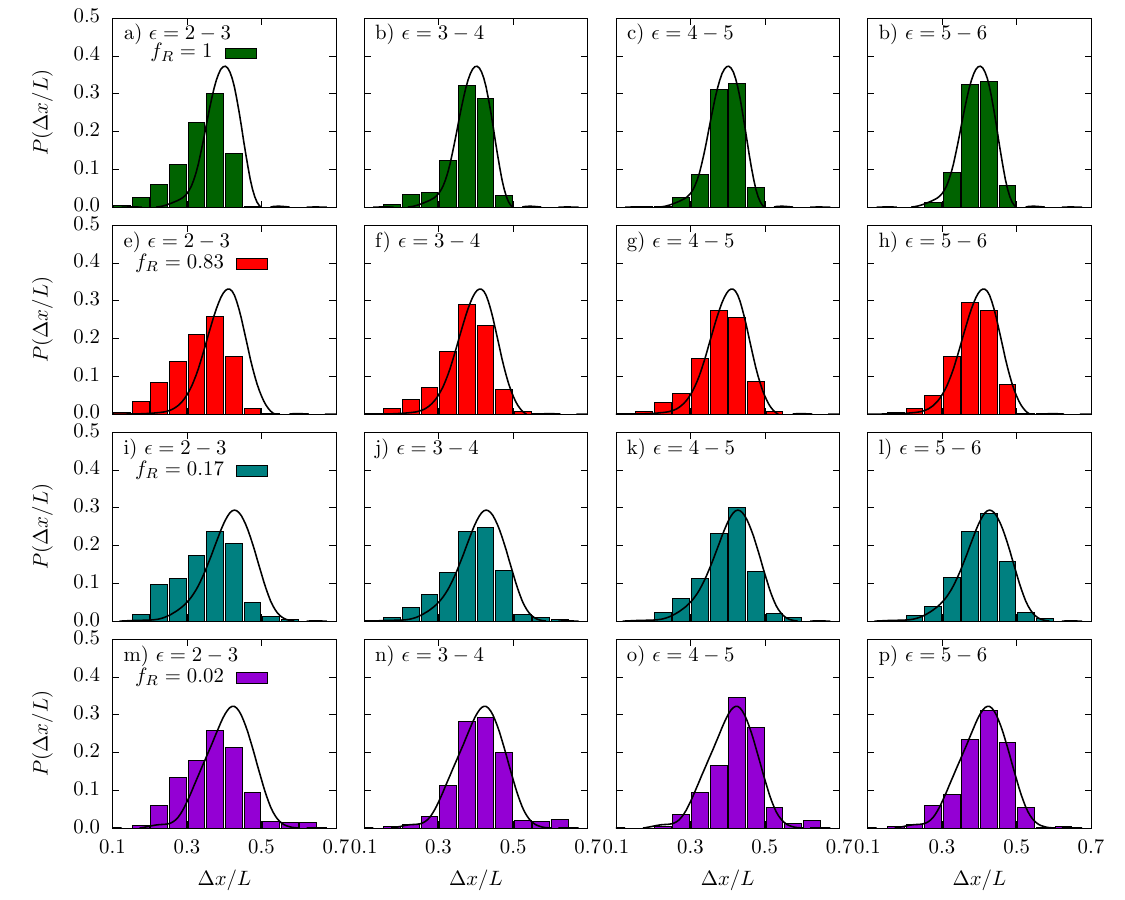}
    \caption{Transient distributions of ring polymer fractional extension for constant $\textrm{Wi}_R \approx 1.6$, increasing Hencky strain ($\epsilon = 2-6$ left to right), and decreasing blend fraction of rings (top to bottom). Blend fractions shown are $f_R = 1$ (a-d), $f_R = 0.83$ (e-h), $f_R = 0.17$ (i-l), and $f_R = 0.02$ (m-p). Rings with a fractional extension $x/L \approx 0.55-0.65$ are threaded with linear chains. Steady state distributions are superimposed as lines.
    }
    \label{fig:dt}
\end{figure*}

We also quantify molecular individualism upon startup flow transience in Fig. \ref{fig:dt}. In particular, we consider a constant strain rate $\textrm{Wi}_R \approx 1.6$ and varying blend ratio at increasing accumulated strain from $\epsilon = 2$ to $\epsilon = 6$. We co-plot instantaneous distributions with the steady state averages from Fig. 6 on the main text, which show transient ring distributions match the steady state conformations after $\epsilon \approx 5$, consistent with the ensemble average extension plateau. First we note features present in all blend ratios; at low strain $\epsilon = 2-4$, distributions are broad as some rings have already stretched to the steady state ensemble average extension, while others remain coiled. This is consistent with previous single molecule experiments and simulations, in which the rate of polymer stretching can be divided into subpopulations based on initial conformations. \cite{li2015ends} As further strain is accumulated, the distributions narrow towards the steady state case as the majority of polymers become stretched.

Transient distributions of ring-linear polymer blends are broader than those of pure ring solutions, similar to the steady state results. This is apparent even at small blend ratios of linear chains, $f_R = 0.83$. As the blend ratio of rings decreases further, the distribution broadens similar to the steady state case. However, we find another feature in startup of flow at $f_R = 0.17,0.02$ which is absent at steady state. There is a small population of rings which are highly stretched to $\Delta x / L \approx 0.55 - 0.65$ for $\epsilon = 2-5$. In a majority blend of linear chains ($f_R = 0.17$) this effect is nearly negligible, but for a single ring in a linear semidilute background ($f_R = 0.02$), the population is more pronounced. The transient distribution even appears shifted to the right of the steady state at $\epsilon = 4-5, f_R = 0.02$. In Section IV of the main text, we show that this occurs due to hooking of linear polymers with rings.

\section{Observation of ring-ring hooks}

\begin{figure*}[htb]
	\includegraphics[width=\textwidth]{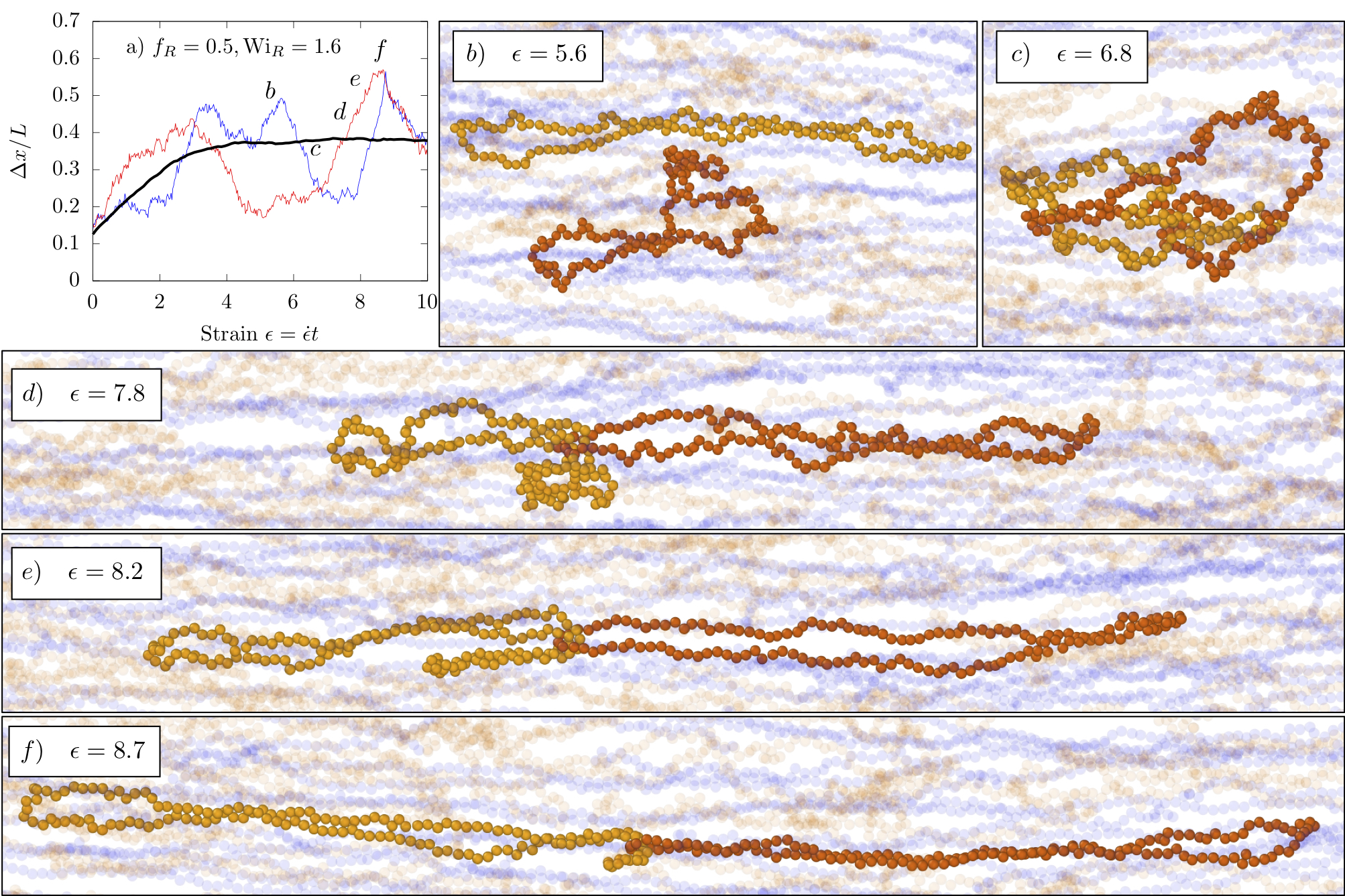}
    \caption{a) Transient fractional extension for two ring polymers forming an intermolecular hook. The position of one ring is fixed at the stagnation point (blue line, light orange in snapshots), and the other is freely advecting (red line, dark orange in snapshots) b) An incompletely stretched ring polymer approaches the stretched ring fixed at the stagnation point c) The freely advecting ring loops around the fixed ring, causing the fixed ring to recoil d) The pair forms a hook compressing the fixed ring in the flow direction e) The freely advecting ring stretches beyond the ensemble average extension as the hook becomes stronger f) Both rings stretch beyond the average extension due the tension of the hook, followed by constraint release and relaxation.
    }
    \label{fig:rrhook}
\end{figure*}

Although ring-ring hooking is rare, we can make some general comments and provide an example in Fig \ref{fig:rrhook}. A weakly stretched ring polymer advects towards a stretched ring polymer with position fixed at the stagnation point (dark-orange and light-orange rings respectively in Fig \ref{fig:rrhook}b). The unstretched advecting ring then loops around the stretched ring and pulls it towards a coiled conformation (Fig \ref{fig:rrhook}c,d). As strain accumulates, the constraint tightens and the ring fixed at the stagnation point adopts a hooked conformation as the advecting ring stretches beyond the ensemble average extension (Fig \ref{fig:rrhook}e). Finally, both rings stretch beyond the ensemble average as the constraint is released (Fig \ref{fig:rrhook}f), followed by relaxation to average levels of extension in both rings.

As opposed to ring-linear hooks, we find ring-ring hooks form between rings which are initially unthreaded at equilibrium. After the ring ensemble average extension plateaus at $\epsilon \approx 4-5$, rings can collide and form strong topological constraints which are not encountered at equilibrium. The excluded volume forces associated with ring-ring hooks are considerably weaker than ring-linear hooks. The tension in the hook is related to the contour length via the flow gradient across the span of the `hooking' polymer. Therefore, at half the contour length of the linear chains, the `hooking' rings impose a weaker constraint and the extra stretching of the `hooked' ring is small. It is not necessary that one of the rings is fixed at the stagnation point, but we speculate that in the case of weak topological constraints it may increase the frequency of intermolecular hooks.

\section{Time dependence of flow modification}

\begin{figure*}[htb]
	\includegraphics[width=\textwidth]{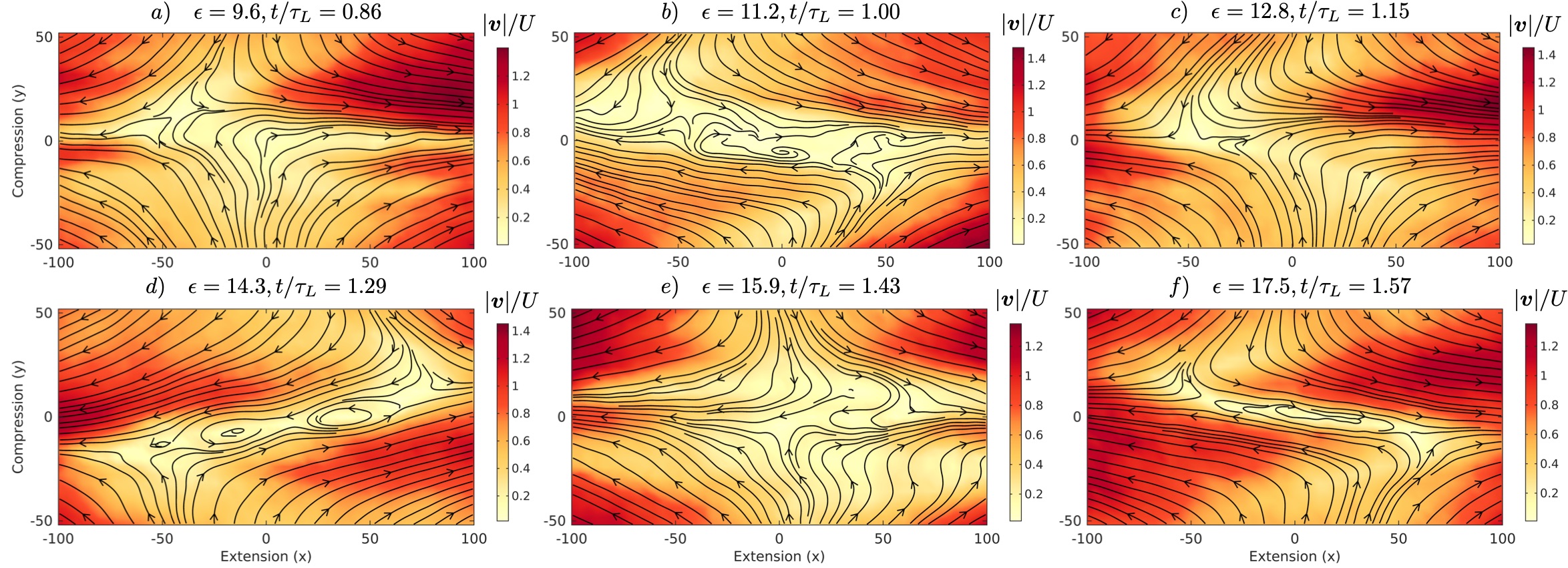}
    \caption{Time sequence of streamlines at $c^*$, $f_R = 0.02$. Fig. 12c is panel d. The flow fluctuates between planar extension (e), stagnation point displacement (a,c), and flow asymmetry (b,d,f).
    }
    \label{fig:t_flow}
\end{figure*}

The deviations from planar extensional flow presented in Fig. 12 of the main text fluctuate significantly in time. In Fig. \ref{fig:t_flow} we show snapshots from the time sequence surrounding Fig. 12c, for a nearly pure linear polymer solution $f_R = 0.02$ at $c^*$ and $\textrm{Wi}_L \approx 11.2$. We find the flow varies between primarily planar extensional (e), stagnation point displacement (a,c), and flow asymmetry across the axis of extension (b,d,f). These fluctuations occur on a time scale significantly faster than the linear polymer relaxation time, and the flow modification is local to the stagnation point. For larger length scales approaching the periodic simulation cell size, the magnitude of disturbance relative to the applied flow is small.

\bibliography{main}